\documentclass[a4paper,11pt]{article}

\pdfoutput=1
\usepackage{jcappub}

\usepackage{graphicx}
\usepackage{longtable}
\usepackage{float}
\usepackage{dcolumn}
\usepackage{bm}
\usepackage{appendix}
\usepackage{multirow}
\usepackage{color}
\usepackage[utf8]{inputenc}
\usepackage{footmisc}
\usepackage{hyperref}

\usepackage{xcolor}
\usepackage{graphicx}
\usepackage{bm}
\usepackage{ulem}
\usepackage{multirow, array, float, tabularx}
\usepackage{amsmath}
\hypersetup{
	citecolor = red,
	linkcolor = blue,
	urlcolor = magenta
}

\newcommand{\be}{\begin{equation}}
	\newcommand{\ee}{\end{equation}}
\newcommand{\bea}{\begin{eqnarray}}
	\newcommand{\eea}{\end{eqnarray}}
\newcommand{\bes}{\begin{subequations}}
	\newcommand{\ees}{\end{subequations}}

\newcommand{\bc}{\begin{center}}
	\newcommand{\ec}{\end{center}}

\usepackage{wasysym}

\begin{document}

    \title{Is natural inflation in agreement with CMB data?}

    \author[a]{F. B. M. dos Santos}\emailAdd{fbmsantos@on.br}
    \author[a]{G. Rodrigues}\emailAdd{gabrielrodrigues@on.br}
    \author[a]{J. G. Rodrigues} \emailAdd{jamersoncg@gmail.com}
    \author[a]{R. de Souza} \emailAdd{souzarayff@gmail.com}
    \author[a]{J. S. Alcaniz}\emailAdd{alcaniz@on.br}

\affiliation[a]{Departamento de Astronomia, Observatório Nacional, 20921-400, Rio de Janeiro - RJ, Brasil}
	
\abstract{Natural inflation is a well-motivated model for the early universe in which an inflaton potential of the pseudo-Nambu-Goldstone form, $V(\phi) = \Lambda^4[1 + \cos{(\phi/f)}]$, can naturally drive a cosmic accelerated epoch. This paper investigates the observational viability of the minimally and non-minimally coupled natural inflation scenarios in light of current Cosmic Microwave Background (CMB) observations. We find that a small and negative coupling of the field with gravity can alleviate the well-known observational discrepancies of the minimally coupled model. We perform a Monte Carlo Markov Chain analysis of the Planck 2018 CMB and BICEP/Keck Array B-mode polarization data to estimate how strong the coupling $\xi$ should be to achieve concordance with data. We also briefly discuss the impact of these results on the physical interpretation of the natural inflation scenario.}

	\maketitle

\section{Introduction}\label{sec1}

In recent years, a great effort has been made to determine what kind of theoretical construction can adequately describe the primordial inflationary era \cite{Starobinsky:1980te,Guth:1980zm,Linde:1981mu,Albrecht:1982wi}. On the other hand, missions that seek to observe and investigate the physical properties of the Cosmic Microwave Background (CMB) have provided valuable insights into the physics of inflation and tight constraints on primordial parameters, such as the amplitude of scalar fluctuations $A_s$ and the spectral index $n_s$. The latest Planck collaboration report \cite{Planck:2018jri,Planck:2018nkj,Planck:2018vyg} shows important results on the status of inflation, either on the primordial power spectrum's behavior or the possible shape of the potential function. Moreover, it allows a direct comparison between many possible scenarios proposed in the literature, and as the constraints on inflationary parameters become more stringent, models that were once viable become discarded. In this context, we also note that extensions of the standard inflationary picture for a particular model can significantly affect the estimates of the inflationary parameters and the CMB theoretical predictions.

One of these extensions is the possibility that the inflaton field is non-minimally coupled to gravity \cite{Futamase:1987ua,Fakir:1990eg,Faraoni:1996rf,Tsujikawa:2004my,Nozari:2007eq,Bauer:2008zj,Okada:2010jf,Hertzberg:2010dc,Tenkanen:2017jih,Rodrigues:2020dod,Rodrigues:2021txa,Campista:2017ovq,dosSantos:2021vis,Santos:2023hhk}, particularly investigated in the context of chaotic inflation. Such construction has gained more attention, since the proposal that a non-minimally coupled Higgs field can drive inflation \cite{Bezrukov:2007ep}. The main attractive feature is that not only is the Higgs field detected, making it possible to link the model's predictions with measured quantities, but it is also possible to achieve a complete inflationary picture without adding exotic new ingredients, except the presence of the non-minimal coupling. Consequently, while the usual minimally coupled Higgs potential would not lead to satisfactory predictions according to CMB data, a non-minimal coupling is responsible for restoring the concordance with the limits on the inflationary parameters.  

In this work, we extend this discussion by investigating another well-motivated construction for the early Universe, namely natural inflation \cite{Freese:1990rb}. This scenario arises when the inflaton is interpreted as a pseudo Nambu-Goldstone boson that appears when a global symmetry is spontaneously broken at a mass scale $f$. As a result, the potential that characterizes the model is shift symmetric in that it is protected from radiative corrections, preserving its flatness during inflation. At a time when most inflationary models were having difficulty in producing results consistent with current data, this model appeared as a viable option, motivated by a solid theoretical background. 

On the other hand, with the increase of more robust observational probes, this model started suffering from drawbacks in its original form. While early Planck and BICEP data showed consistency with natural inflation at the level of parameters $n_s$ and $r$ (the scalar spectral index and tensor-to-scalar ratio, respectively), the following reports started to show a tendency for exclusion of the model. This trend continued in the latest Planck 2018 report \cite{Planck:2018jri}, in which the model is highly disfavored by data, especially when the BICEP/Keck Array data is considered into the analysis. In particular, these later data proved invaluable for determining viable inflationary models and the possibility of detecting gravitational waves (GWs) produced during inflation. Currently, the best experiments dedicated to the investigation of CMB B-modes are the BICEP/Keck Array telescopes, located at the south pole \cite{BICEP1:2013sbv,BICEP2:2018kqh,BICEP:2021xfz}, with the latest observation season report bringing some striking results regarding the constraints on potential inflationary models. The most recent data from the 2018 observing season (BK18) \cite{BICEP:2021xfz} show how the $n_s-r$ parameters are constrained for the standard $\Lambda$CDM model. Compared with simple inflationary models (see Fig. 5 of \cite{BICEP:2021xfz}), we note that the natural inflation model is entirely excluded by data when Planck+lensing+BK18+BAO data are considered.  

While a generalization of the original potential can be a way out of this strong tension with data \cite{Gerbino:2016sgw}, the possibility that the scalar field is non-minimally coupled to gravity has recently gained some attention. The first approach was to assume that the coupling has the same properties of the potential, that is, being shift symmetric \cite{Ferreira:2018nav}. An analysis at the inflationary parameters level was conducted, and a noticeable alleviation was perceived. However, the question on whether natural inflation with a non-minimal coupling is consistent with the latest CMB data still needs a clear answer. 

In our discussion, we seek to answer this question by testing the observational viability of the non-minimally coupled natural inflation model, considering the simplest and well-motivated coupling $\xi R\phi^2$~\cite{Reyimuaji:2020goi}. In this later work, the authors found that the tension with data can be solved if a negative non-minimal coupling, characterized by the constant $\xi$, is considered and the concordance with the $n_s-r$ confidence contours can be restored. A discussion with Planck+BK18 data was presented in \cite{Bostan:2022swq,Bostan:2023ped}. However, a complete statistical analysis is still lacking. In our analysis, we apply the Bayesian parameter selection procedure to obtain the magnitude of the non-minimal coupling $\xi$, the non-perturbative energy scale $\Lambda$, and the scale of the symmetry breaking $f$ favored by current CMB data. We also apply statistical criteria to compare the minimal/non-minimal natural inflation model with the canonical $\Lambda$CDM scenario.

This work is organized as follows. In Section \ref{sec:2}, we review the original natural inflation model and its non-minimally coupled extension. Section \ref{sec:3} is devoted to the methodology and data used in our subsequent numerical analysis, while in Sec. \ref{sec:4} we discuss our results. In Sec. \ref{sec:5} we present our main conclusions and perspectives.

\section{Natural inflation}\label{sec:2}

In this section, we present the idea of natural inflation in more detail, focusing on the main consequences of a non-minimally coupled scalar field within the model.

\subsection{The original model}

The natural inflation scenario \cite{Freese:1990rb} arises from the assumption that inflation is driven by an axion-like particle (ALP). Such class of fields arise naturally in the context of string theory and supergravity, where they are associated to the geometry of compact spatial dimensions~\cite{Kim:1986ax,Svrcek:2006yi}. In the context of field theory, the inflaton is introduced as a pseudo-Goldstone boson resulting from the spontaneous breaking of a global $U(1)$ symmetry. A non-trivial vacuum structure, generally associated to non-perturbative effects, is generated at the energy scale $\Lambda$. The resulting potential energy takes the form,  
\begin{equation}
V(\phi) = \Lambda^4\left[1+\cos\left(\frac{\phi}{f}\right)\right], \label{eq:2.1}
\end{equation}
with $f$ characterizing the $U(1)$ symmetry breaking scale.

In order to obtain the conditions for inflation, one shall find the parameter space region where the aforementioned scalar potential assumes an approximately flat behavior for a sufficiently large region in field space. In the minimally coupled natural inflation, this is obtained for large values for $\Lambda$ and $f$, namely $\Lambda \sim 10^{16}$ GeV and $f \gtrsim M_P$, where $M_P$ is the reduced Planck mass, defined as $M_P=\frac{1}{\sqrt{8\pi G}}$. Such configuration prevents the canonical QCD axion to perform the role of the inflaton, once $\Lambda_{QCD} \sim 300$ MeV is dictated by the non-perturbative behavior of the quark fields \cite{Peccei:1977hh,Wilczek:1977pj,Weinberg:1977ma,Kim:1979if,Shifman:1979if,Dine:1981rt,Zhitnitsky:1980tq}. The inflationary parameters are obtained following the usual slow-roll parameters,
\begin{eqnarray}
 &\epsilon & = \frac{M^2_{P}}{2}\left(\frac{ V^{\prime}}{ V}\right)^2, \quad \quad
 \eta  = M^2_{P}\left( \frac{V^{\prime \prime}}{V } \right),
 \label{eq:2.2}
\end{eqnarray}
where $^\prime$ indicates derivative with respect to the inflaton. In the slow-roll approximation, one can compute the predictions for the spectral index and the tensor-to-scalar ratio through,
\begin{equation}
 n_s=1-6\epsilon_\star+2\eta_\star, \quad \quad \quad r=16\epsilon_\star,
 \label{eq:2.3}
\end{equation}
with the subscript $_\star$ denoting that these quantities are measured when the pivot scale, $k_\star$, leaves the horizon. The Planck Collaboration \cite{Planck:2018jri} estimates $n_s=0.9651\pm0.0041$ at $68\%$ confidence level (C.L.). When combined with the  BICEP/Keck collaboration (BK) data \cite{BICEP2:2018kqh}, the tensor-to-scalar ratio is constrained as $r<0.056$ (95\% C.L.), while more recent results from the 2018 observation season (BK18) \cite{BICEP:2021xfz} tighten the constraint to $r<0.035$, when Planck+BK18 is added to baryonic acoustic oscillations (BAO) data.

The amplitude of the primordial spectrum, defined as $\Delta^2_{\mathcal{R},\star}=A_s$, assumes the form under the slow-roll approximations,
\begin{equation}
	A_s=\frac{V_\star}{24\pi^2\epsilon_\star M_P^4}\Big|_{k_\star}\;,
    \label{eq:2.4}
\end{equation}
The Planck collaboration estimates $\log_{10}\left(10^{10}A_s\right)=3.044\pm 0.014$, also at $68\%$ (C.L.). Its connection with the inflationary potential allows us to estimate one of the free parameters of the model through $A_s$, precisely the scale $\Lambda$, consequently providing an extra constraint on the specific model when confronted with cosmological data.

\begin{figure*}
    \centering
	\includegraphics[width=0.7\columnwidth]{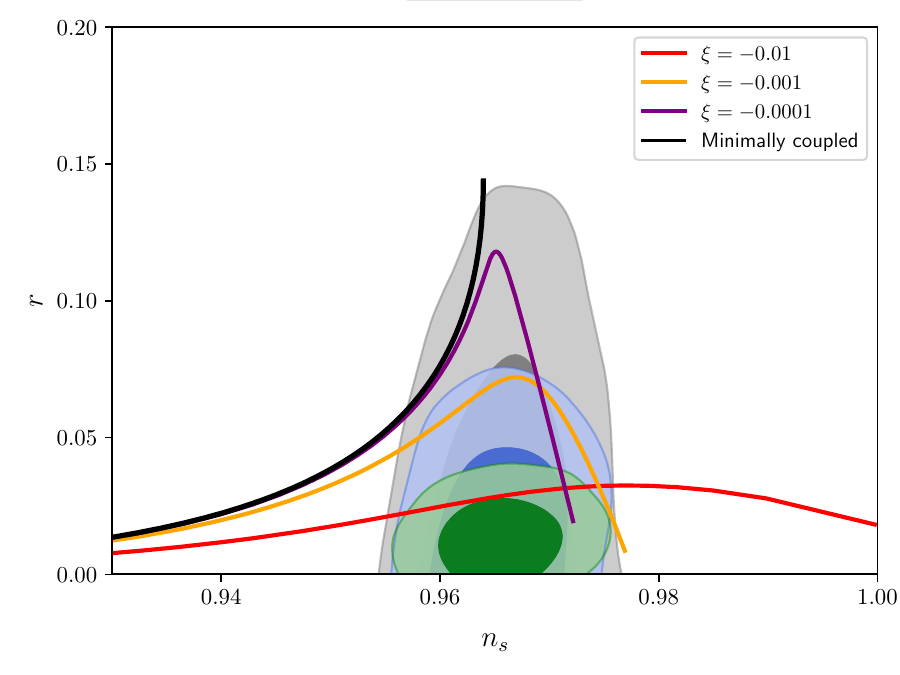}
	\caption{The $n_s$ and $r$ predictions for the minimally (black curve) and non-minimally coupled models, considering $N_\star=55$, plotted against the contours on the inflationary parameters given by Planck TTTEEE+lowl+lensing (gray), Planck TTTEEE+lowl+lensing+BK15 (blue), and Planck TTTEEE+lowl+lensing+BK18 (green).}
	\label{fig:1}
\end{figure*}

In Fig.~\ref{fig:1}, we show the predictions of the minimally coupled natural inflation model in the $n_s\, \times \, r$ plane (black curve). The values of $f/M_p$ increase as the curve goes to higher $r$. We note a possible disagreement with data, when considering the latest Planck results. For Planck+TTTEEE+lowl+lensing, there is a better agreement within the $2\sigma$ confidence region; on the other hand, when the B-mode polarization data from the BICEP/Keck Array collaboration is added, the minimally coupled model is basically excluded, which is very noticeable for the Planck+BK18 constraints (green contours). This situation shows that to make natural inflation a viable scenario, some extension of the standard cosmological theory must be considered.

\subsection{Non-minimally coupled natural inflation}

It is argued that terms that involve a direct coupling between a scalar field and the geometric sector are unavoidable in the context of quantum field theory in curved space-time~\cite{Birrell:1982ix}, in a manner that the strength of this coupling might have important implications in the description of cosmological dynamics. By considering the following action for a scalar field $\phi$
\begin{equation}\label{eq:2.5}
S = \int d^4x\sqrt{-g}\left[\frac{M_P^2}{2}\Omega^2(\phi)R + \frac{1}{2}g^{\mu\nu}\partial_\mu\phi\partial_\nu\phi - V_J(\phi)\right],
\end{equation}
with $\Omega^2(\phi)$ being the general term that represents the non-minimal coupling, we can obtain the dynamics of the field from a specific potential $V_J(\phi)$. While eq. (\ref{eq:2.5}) characterizes the so-called Jordan frame, in order to better implement our subsequent numerical analysis, we shall work on the Einstein frame, characterized by the conformal transformation of the metric $\tilde g_{\mu\nu}=\Omega^2g_{\mu\nu}$~\cite{Accioly:1993kc,Faraoni:1998qx}. This allows us to rewrite the action above as
\begin{equation}\label{eq:2.6}
S = \int d^4x\sqrt{-\tilde g}\left[\frac{M_P^2}{2}\tilde R +  \frac{1}{2}\tilde g^{\mu\nu}\partial_\mu\chi\partial_\nu\chi - V_E(\chi)\right],
\end{equation}
written in terms of a new field $\chi$, which relates to the Jordan frame field $\phi$ through the conformal function $\Omega^2(\phi)$ as
\begin{equation}
    \frac{d\chi}{d\phi}=\sqrt{\frac{\Omega^2 +(3/2)M_P^2(d\Omega^2/d\phi)^2}{\Omega^4}}.\label{eq:2.7}
\end{equation}
Note that the potential is also redefined as $V_E(\chi(\phi))\equiv\frac{V_J(\phi)}{\Omega^4}$, encapsulating the extra complexity given by the non-minimal coupling. In recent literature, it has been argued that the presence of a very weak non-minimal coupling to gravity might alleviate the tension between the predictions of inflationary parameters given by the natural inflation model \cite{Reyimuaji:2020goi}. Although the form of the coupling function is not unique, in this work we consider the following form discussed in \cite{Reyimuaji:2020goi}

\begin{equation}
\Omega^2(\phi)=1+\xi\left(\frac{\phi}{M_P}\right)^2,
\label{eq:2.8}
\end{equation}
which is well-motivated theoretically, being the most used in the studies involving the non-minimal coupling of a scalar field with gravity. 

\begin{figure}
\centering
\includegraphics[width=0.485\textwidth]{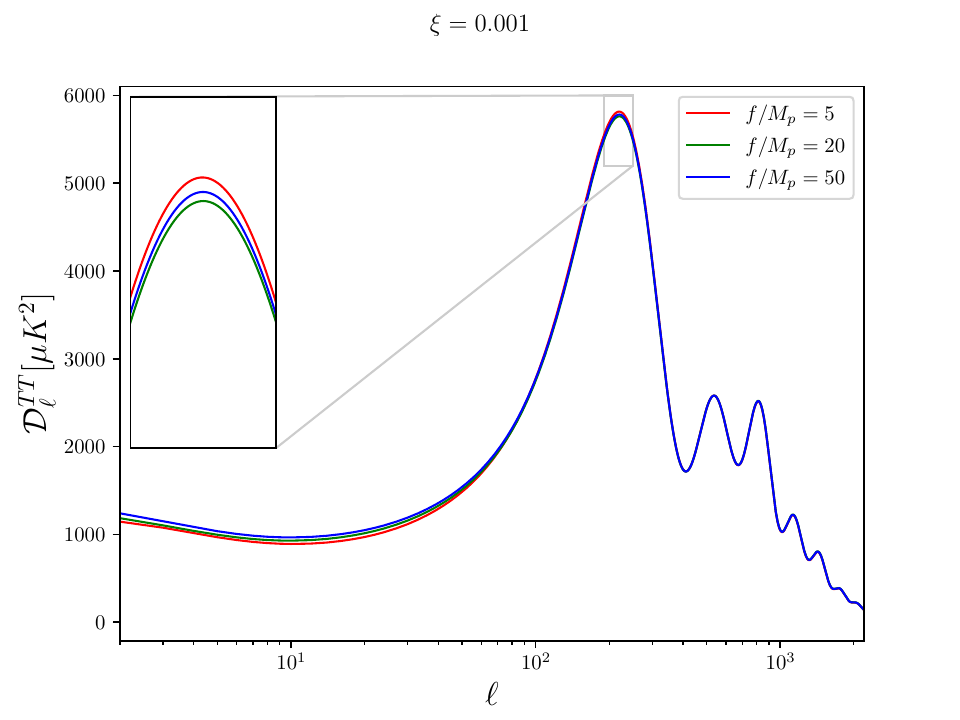}
\includegraphics[width=0.485\textwidth]{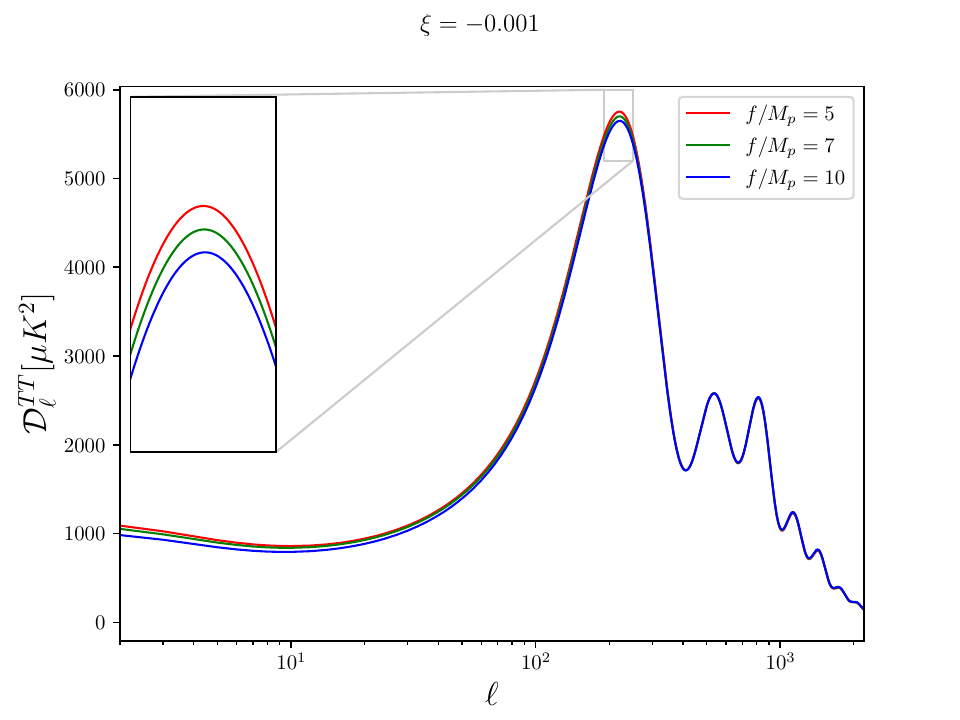}\\
\includegraphics[width=0.453\textwidth]{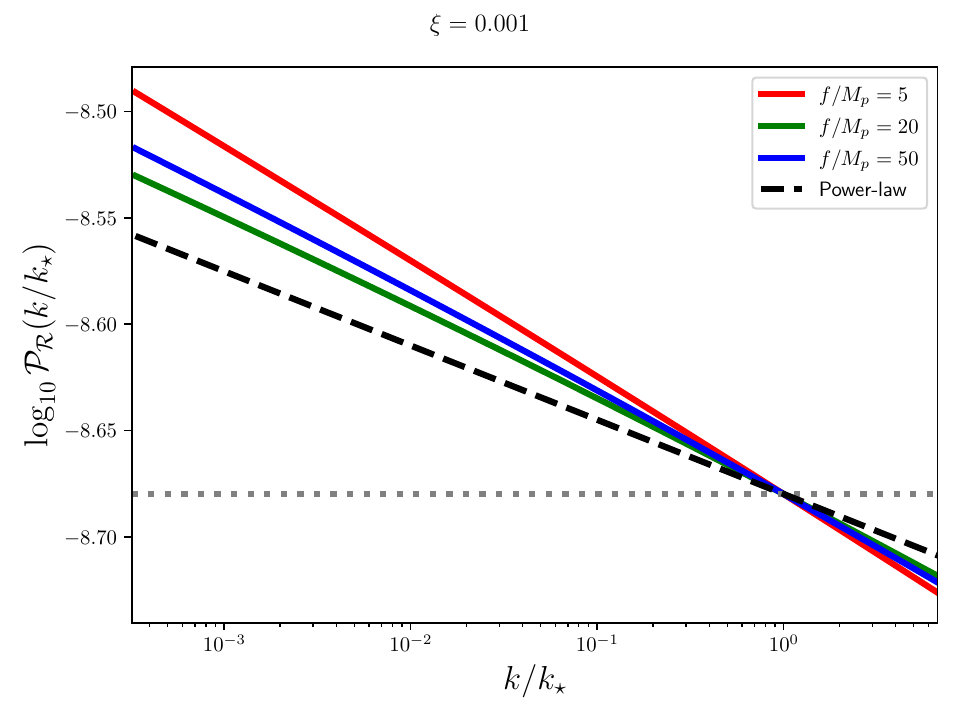}
\hspace{0.53cm}\includegraphics[width=0.455\textwidth]{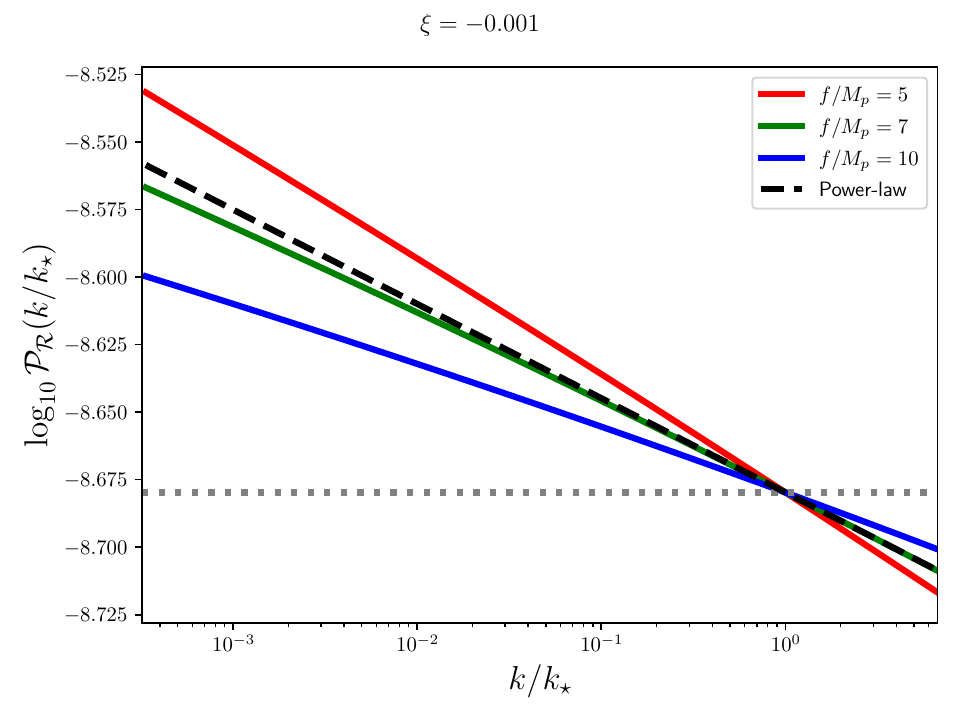}\hspace{0.35cm}
\caption{The CMB temperature power spectra (upper panels) and the primordial spectra of the scalar perturbations (lower panels) for the non-minimally coupled natural inflation scenario. We fix $\xi=0.001$ (left), and $\xi=-0.001$ (right) and select specific values for $f/M_P$. The dotted gray line in the lower panels indicate the amplitude of the spectrum at the pivot scale, which is fixed, so all curves will converge at that point.}\label{fig:2}
\end{figure}

Let us now review the main results of non-minimally coupled natural inflation obtained in recent literature. In \cite{Ferreira:2018nav}, a coupling of periodic form was considered, motivated by the shift symmetry of the potential. The authors argued that the function $\Omega^2(\phi)=1+\xi(1+\cos\phi/f)$ is the simplest one that presents this feature, and performed the slow-roll analysis of the scenario. For a positive coupling $\xi$, some discrepancies with the Planck constraints on the $n_s$ and $r$ parameters are alleviated, as well as the issue of the magnitude of the periodicity scale $f$, allowed to have values closer to $M_P$ with respect to the minimally coupled case. Interestingly enough, a stronger coupling of the field with gravity is allowed, up to $\xi\sim 20$, being a threshold value for which higher ones would essentially lead to the same curves in the $n_s-r$ parameter space. One might note that this scenario can be closely related to the one proposed in \cite{Kallosh:2013tua}, in which one can achieve an attractor behavior in the $n_s-r$ plane if the potential and coupling are related as $V(\phi)=\Lambda^4 f^2(\phi)$ and $\Omega^2(\phi)=1+\xi f(\phi)$, where $f(\phi)$ is a general function; this means that for any model, in the strong coupling limit, the predictions of the inflationary parameters will go toward the well-known ones predicted by the Starobinsky and $\alpha$-attractor models in the $\alpha\rightarrow 0$ limit \cite{Kallosh:2013hoa,Kallosh:2013tua,Kallosh:2013yoa}.

The impact of a nonzero $\xi$ in the predictions of the model for the coupling given by eq. (\ref{eq:2.8}) was recently well discussed in \cite{Reyimuaji:2020goi}, where both metric and Palatini formalisms were investigated. In general, both formalisms yield almost the same results at the level of inflationary parameters; this is due to the fact that a very small non-minimal coupling of the field is favored by the $n_s$ and $r$ restrictions, being of order $\xi\sim 10^{-4}-10^{-3}$ (see also \cite{Campista:2017ovq,dosSantos:2021vis,Santos:2023hhk}, for other models where that is also the case). The most striking aspect is the significant impact that such a small coupling can realize in the slow-roll predictions of the model, being able to restore the concordance of the inflationary parameters with data. Fig. \ref{fig:1} also shows the predictions for $n_s$ and $r$ for the extended model. We consider fixed values of $\xi$, while varying the mass scale $f/M_P$ respecting the upper limit $\frac{f}{M_P} < \frac{1}{\sqrt{|\xi|}\pi}$ (for a negative $\xi$), necessary for the potential to be well-behaved and support the slow-roll regime. We first note that, while for small negative values (e.g., $\xi=-0.0001$), the curve sharply goes toward lower $r$ as $f/M_P$ increases, for more negative $\xi$, we notice a tendency for larger values of the spectral index for an increasing $f/M_P$, as represented by the red curve, where $\xi=-0.01$.

Figure \ref{fig:2} shows the CMB temperature power spectra (upper panels) and the primordial spectrum of scalar perturbations (lower panels) for the non-minimally coupled natural inflation model. We have fixed the $\xi$ parameter while varying $f$ to notice how differently the curves behave as a description of the CMB temperature anisotropies. By setting $\xi=0.001$, as shown on the upper left panel, we see that as $f$ increases, the amplitude of the spectra in its first peak change in a non-trivial manner, since we do not see a direct relation between the increase or decrease of the spectrum and the change in $f/M_P$. This relation looks a bit more clear when we choose $\xi=-0.001$ (upper right panel); we note that as $f/M_P$ increases, the amplitude of the temperature spectrum seems to decrease in the first peak. The same qualitative behavior can be observed in the primordial scalar perturbations (bottom panels), where the different slopes of the lines are associated with different values of the spectral index, $n_S$.  This kind of behavior suggests that only the full analysis will be able to tell what kind of correlation exist between the two parameters in the description of the data. The main correlation expected will most certainly be imposed by the condition $\frac{f}{M_p} < \frac{1}{\sqrt{|\xi|}\pi}$, which sets the maximum value possible for the symmetry-breaking scale for a given $\xi$. In the following sections, we will show that this has important consequences on the constraints for these two parameters and the ability of the model to solve tensions within natural inflation.

\section{Methodology and data sets}\label{sec:3}

As the impact on the inflationary parameters is very significant when $\xi\neq 0$, it is worth investigating the general impact of the extended scenario in the description of the anisotropies of the CMB. Discrepancies between the results of the $n_s-r$ plane and the best fit of the power spectra when the non-minimally coupled models are compared to the Planck data have been found and discussed before \cite{Campista:2017ovq}. Therefore, to determine the possible viability of a specific scenario, one must seek to perform a complete numerical analysis that can constrain the parameter space while also guaranteeing consistency with all data \cite{Giare:2023kiv}. 

\begin{figure}
\centering
\includegraphics[width=0.5\textwidth]{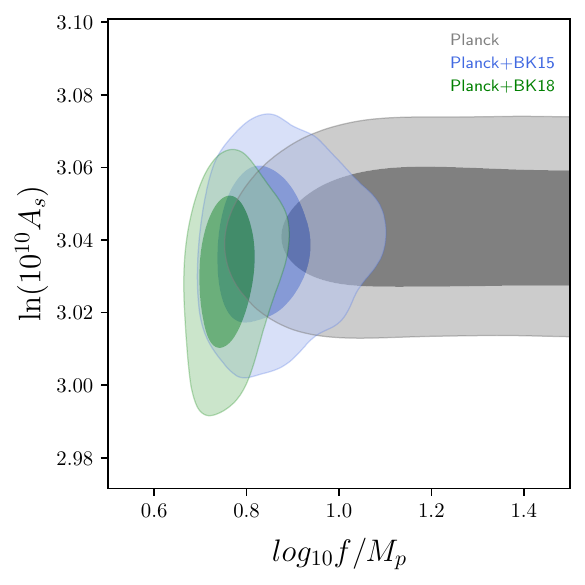}
\caption{Confidence contours at 68\% and 95\% C.L. for the minimally coupled natural inflation model. We show the correlation between $\ln(10^{10}A_s)$ and $\log_{10}f/M_p$, when Planck (gray), Planck+BK15 (blue) and Planck+BK18 (green) data are considered in the analyses.}\label{fig:3}
\end{figure}

To perform such analysis, we implement the model in the Cosmic Linear Anisotropy Solving System (CLASS) \cite{Lesgourgues:2011re,Blas:2011rf}, which allows us to produce the spectra by solving the Klein-Gordon and Mukhanov-Sasaki equations, for the curvature perturbation, found as $\mathcal{P}_\mathcal{R}=\frac{k^3}{2\pi^2}\left|\frac{u_k}{z}\right|^2$, given a proper parametrization of the model. In our analysis, together with the CLASS code, we interface it with MontePython \cite{Brinckmann:2018cvx,Audren:2012wb}, which uses the CLASS output as an input for performing the statistical analysis. Specifically, we perform a Monte Carlo Markov Chain (MCMC) approach using the Metropolis-Hastings method. This is done by varying the usual cosmological parameters, being the baryon and cold dark matter density parameters $\Omega_bh^2,\Omega_ch^2$, the ratio of the sound horizon and the angular distance $\theta$ at decoupling, the optical depth $\tau$ and the amplitude of the primordial spectrum $A_s$. For the standard $\Lambda$CDM, the spectral index $n_s$ is the sixth parameter that characterizes the model, from which we add the tensor-to-scalar ratio $r$ as a varying parameter, while in the inflationary model considered, both are derived parameters. We then vary the model parameters $\xi$ and $\text{log}_{10}f/M_P$ freely, as allowed by the MontePython code, but leave an upper limit on $f$ as $\log_{10}f/M_P=2.5$ \footnote{Higher values of $f/M_p$ could violate the condition $f/M_P<1/(\sqrt{\xi}\pi)$ when negative $\xi$ are considered, as mentioned earlier in the text; we have taken care of this issue in the analysis by excluding the possibility of such points of being chosen within the code, such that only the well-behaved points are considered.}. The minimally coupled model, when $\xi=0$ is denoted in the results as MCNI, while the non-minimally coupled one is denoted as NMCNI. As for the parameter $\Lambda^4$, it is obtained through eq. (\ref{eq:2.4}), therefore establishing a relation between this parameter and $\xi$, $f$ and $A_s$. We also fix the sum of neutrino masses to $0.06$ eV and use the pivot scale set by Planck, of $k_\star=0.05$ Mpc$^{-1}$, for the primordial spectrum, while $r$ is computed at $k_\star=0.002$ Mpc$^{-1}$. Finally, we also choose $N_\star=55$, as the number of e-folds left until the end of inflation, and the resulting chains will be analysed with the GetDist module \cite{Lewis:2019xzd}.

For the data used, we choose the latest Planck 2018 results \cite{Planck:2018jri} for the TTTEEE+lensing likelihoods, where the low-multipole $\ell$ for both temperature-temperature and EE modes are included. We denote this as the `Planck' data set. Additionally, we also include the B-mode polarization data from the BICEP/Keck collaboration, in our analysis. As shown earlier (see Fig. \ref{fig:1}), these data strongly disfavor the natural inflation model, as they impose an even more severe upper limit on the tensor-to-scalar ratio. In order to track the constraints when these data are considered, we perform the analysis for the last two data releases from the 2015 (BK15) and 2018 (BK18) observation periods, added with Planck. Thus, we expect a more definitive answer on whether a non-minimal coupling is really able to solve the tension between the model and cosmological data. 

An important issue that should be added to this discussion is whether this or any proposed model is preferred by data when compared to a reference model. This is essential, as a specific model could have enough free parameters to fit the cosmological data perfectly. However, according to statistical criteria, the additional complexity plays against the model's validity. Here, we compare both minimally and non-minimally coupled models with the standard $\Lambda$CDM one by using the Deviance Information Criterion (DIC), defined as

\begin{equation}
    \text{DIC} = \overline{D(\theta)} + p_D,
\end{equation}
with $p_D =  \overline{\chi^2(\theta)} - {\chi^2}(\overline\theta)$ being the Bayesian complexity, $\chi^2$ being the chi-squared function whereas $\theta$ characterizes the parameters of a given model being varied. The preference for a model is obtained by using the Jeffreys' scale, i.e., a difference $\Delta$DIC$\equiv$DIC$_{\text{model}}-$DIC$_{\text{ref}}$ $>5$ or $>10$ signals a strong or decisive evidence against the proposed model \cite{Liddle:2007fy}.

\begin{figure}
\centering
\includegraphics[width=\textwidth]{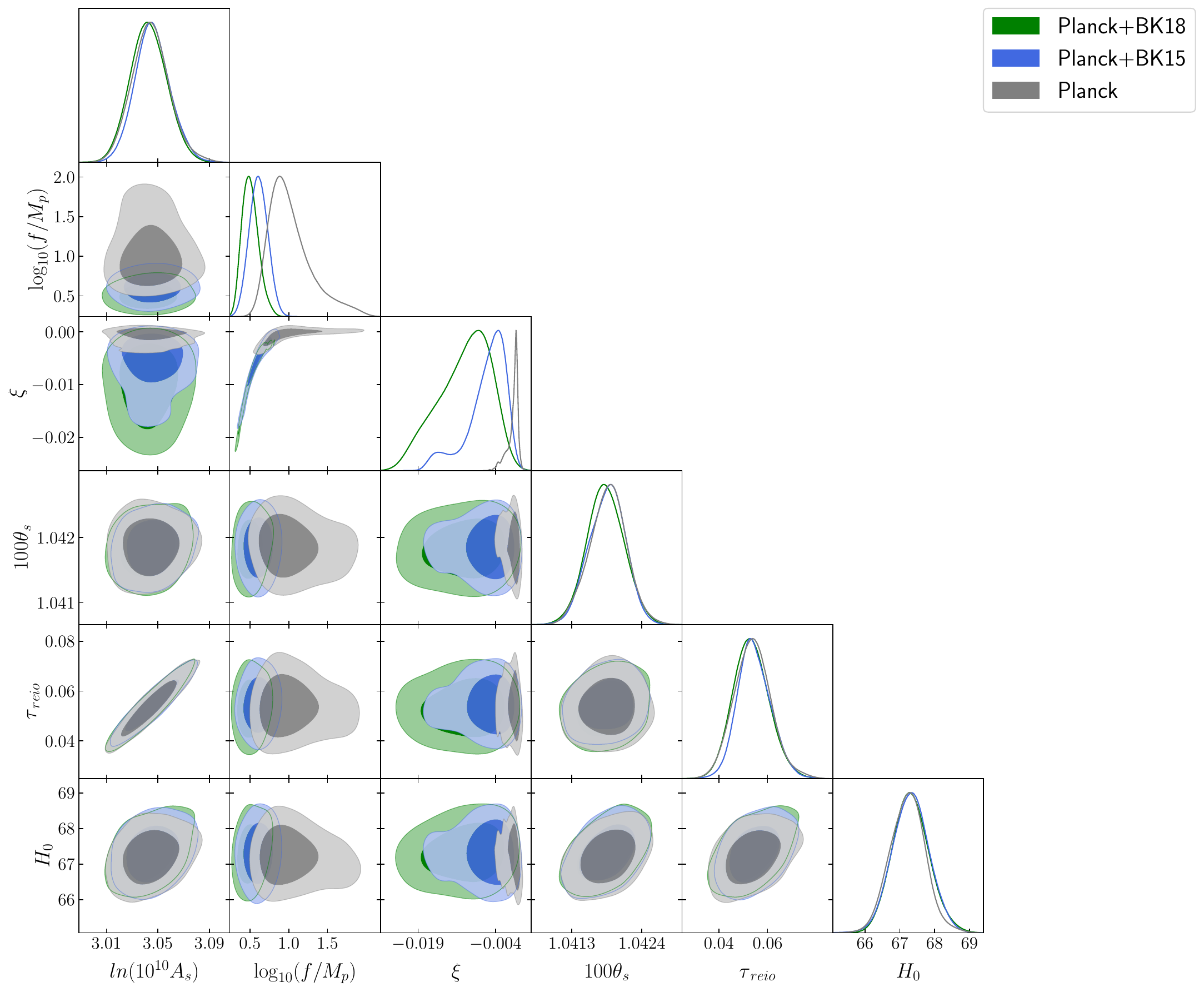}
\caption{Confidence contours at 68\% and 95\% C.L. for the non-minimally coupled natural inflation model. The results are shown when Planck (gray), Planck+BK15 (blue) and Planck+BK18 (green) data are considered in the analyses.}\label{fig:4}
\end{figure}

\section{Results and discussion}\label{sec:4}

Considering B-mode polarization data results in interesting outcomes for minimally and non-minimally coupled models. First, we start with the minimally coupled one (MCNI). The Planck 2015 report \cite{Planck:2015sxf} presented the results for the $\xi=0$ case, for Planck TT+lowP+BAO data sets. They estimated only a lower limit on $\log_{10}f/M_p$, of $\log_{10}f/M_p>0.84$. In our analysis, we obtained the same behavior, for the `Planck' data set, finding $\log_{10}f/M_p>0.91$. The constraints for all models analysed are shown in Table \ref{tab:1}, while the confidence contours for the $\text{ln}(10^{10}A_s)-\text{log}_{10}f/M_p$ plane are shown in Fig. \ref{fig:3}. Adding the BK15 and BK18 data sets leads to a better constraint on $f/M_p$, for which the mass scale is driven towards lower values due to its correlation with the tensor-to-scalar ratio $r$. For `Planck+BK15', we estimated $\text{log}_{10}f/M_p=0.878^{+0.053}_{-0.12}$, with central value corresponding to $f/M_p\simeq 7.55$, which significantly alleviates the superplanckian value problem. However, this does not mean that the concordance with the $n_s-r$ constraints is restored in any manner. As discussed earlier, the results considering B-mode polarization data show the observational inviability of the minimally coupled natural inflation model. For this model, we obtain $n_s=0.9611$ and $r=0.0851$.

\begin{table*}[t]
		\centering
		\begin{tabular}{>{\scriptsize}c >{\scriptsize}c >{\scriptsize}c >{\scriptsize}c >{\scriptsize}c >{\scriptsize}c  >{\scriptsize}c}
			\hline
			\hline
			& $\Lambda$CDM$+r$ & & MCNI & & NMCNI & \\
			\hline
			{Parameter} & {mean} & {best fit} & {mean} & {best fit}& {mean} & {best fit}\\
            \hline 
            \multicolumn{7}{c}{\scriptsize\textbf{Planck}}\\
			
			$100\Omega_b h^2$   & $2.237\pm 0.0147$ & $2.245$     & $2.232\pm 0.014$ & $2.245$  & $2.233\pm 0.014$ & $2.237$ \\
			$\Omega_{c} h^2$ & $0.119\pm 0.0011$ & $0.119$     & $0.12069^{+0.00096}_{-0.0011}$ & $0.120$    & $0.1207\pm 0.0012$ & $0.120$ \\
			$100\theta$ & $1.0418\pm 0.000304$ & $1.0415$     & $1.04180\pm 0.00031$ & $1.0416$    & $1.04180\pm 0.00029$ & $1.0418$ \\
			$\tau$ & $0.0545\pm 0.00752$ & $0.0544$     & $0.0514\pm 0.0077$ & $0.0530$    & $0.0525\pm 0.0078$ & $0.0576$ \\
            $\ln(10^{10}A_s)$ & $3.044\pm 0.0145$ & $3.049$     & $3.041\pm 0.015$ & $3.045$    & $3.043\pm 0.015$ & $3.051$ \\
			$\log_{10}(f/M_p)$ & $-$ & $-$     & $>0.91$ & $1.002$   & $0.974^{+0.097}_{-0.28}$ & $0.702$ \\
			$\xi$ & $-$ & $-$     & $-$ & $-$    & $-0.00036^{+0.00070}_{-0.00014}$ & $-0.00280$ \\
			$n_s$  & $0.9664\pm 0.004$ & $0.9663$  & $-$ & $-$    & $-$ & $-$ \\
			$r$ & $<0.15$ & $0.0036$  & $-$ & $-$    & $-$ & $-$ \\
			$H_{0}^{\ast}$ [Km/s/Mpc] & $67.45\pm 0.53$ & $67.45$  & $67.08^{+0.51}_{-0.44}$ & $67.35$    & $67.10\pm 0.51$ & $67.31$ \\
            \hline
			$\Delta$DIC & Reference & & $-1.384$ & & $-0.833$ & \\
			\hline
   
            \multicolumn{7}{c}{\scriptsize\textbf{Planck + BK15}}\\
			
			$100\Omega_b h^2$   & $2.237\pm 0.014$ & $2.229$     & $2.228\pm 0.014$ & $2.215$    & $2.235\pm 0.015$ & $2.242$ \\
			$\Omega_{c} h^2$ & $0.12\pm 0.0012$ & $0.120$     & $0.1212\pm 0.0010$ & $0.121$    & $0.1201\pm 0.0012$ & $0.120$ \\
			$100\theta$ & $1.04188\pm 0.00030$ & $1.0418$     & $1.04174\pm 0.00030$ & $1.0417$    & $1.04183\pm 0.00028$ & $1.042$ \\
			$\tau$ & $0.0556\pm 0.0076$ & $0.0527$     & $0.0505\pm 0.0069$ & $0.0482$    & $0.0547^{+0.0062}_{-0.0075}$ & $0.0507$ \\
            $\ln(10^{10}A_s)$ & $3.048\pm 0.015$ & $3.047$     & $3.040\pm 0.014$ & $3.032$    & $3.046^{+0.012}_{-0.014}$ & $3.038$ \\
			$\log_{10}(f/M_p)$ & $-$ & $-$     & $0.878^{+0.053}_{-0.12}$ & $0.790$    & $0.61\pm 0.12$ & $0.59$ \\
			$\xi$ & $-$ & $-$  & $-$ & $-$    & $-0.0055^{+0.0044}_{-0.0018}$ & $-0.00526$ \\
			$n_s$  & $0.9658\pm 0.0042$ & $0.9641$  & $-$ & $-$    & $-$ & $-$ \\
			$r$ & $<0.061$ & $0.024$  & $-$ & $-$    & $-$ & $-$ \\
			$H_{0}^{\ast}$ [Km/s/Mpc] & $67.39\pm 0.54$ & $67.16$  & $66.85\pm 0.46$ & $66.49$    & $67.32\pm 0.52$ & $67.36$ \\
            \hline
			$\Delta$DIC & Reference & & $3.951$ & & $-0.184$ & \\
			\hline
            \multicolumn{7}{c}{\scriptsize\textbf{Planck + BK18}}\\
			
			$100\Omega_b h^2$   & $2.238\pm 0.015$ & $2.236$     & $2.225\pm 0.013$ & $2.224$    & $2.234\pm 0.015$ & $2.240$ \\
			$\Omega_{c} h^2$ & $0.12\pm 0.0012$ & $0.120$     & $0.1216\pm 0.0012$ & $0.121$    & $0.1204^{+0.0012}_{-0.0014}$ & $0.120$ \\
			$100\theta$ & $1.04188\pm 0.00029$ & $1.0415$     & $1.04168\pm 0.00029$ & $1.0418$    & $1.04181\pm 0.00030$  & $1.0421$ \\
			$\tau$ & $0.0547^{+0.0068}_{-0.0077}$ & $0.0497$ & $0.0467\pm 0.0073$  & $0.0487$    & $0.0522\pm 0.0078$ & $0.0499$ \\
            $\ln(10^{10}A_s)$ & $3.046\pm 0.014$ & $3.038$     & $3.034^{+0.014}_{-0.013}$ & $3.034$    & $3.042\pm 0.015$ & $3.036$ \\
			$\log_{10}(f/M_p)$ & $-$ & $-$     & $0.793^{+0.026}_{-0.089}$ & $0.757$    & $0.55^{+0.11}_{-0.15}$ & $0.434$ \\
			$\xi$ & $-$ & $-$     & $-$ & $-$    & $-0.0081^{+0.0065}_{-0.0051}$ & $-0.0126$ \\
			$n_s$  & $0.9657\pm 0.0041$ & $0.9664$  & $-$ & $-$    & $-$ & $-$ \\
			$r$ & $<0.0348$ & $0.011$  & $-$ & $-$    & $-$ & $-$ \\
			$H_{0}^{\ast}$ [Km/s/Mpc] & $67.40\pm 0.53$ & $67.12$   & $66.66\pm 0.51$ & $66.90$  & $67.21\pm 0.59$ & $67.38$ \\
            \hline
			$\Delta$DIC & Reference & & $10.662$ & & $2.869$ & \\
			\hline
		\end{tabular}
		\caption{The estimates at $68\%$ confidence level (C.L.) and best-fit values for the cosmological parameters, when Planck TT,TE,EE+lensing (labeled `Planck' - upper), Planck+BK15 (center) and Planck+BK18 (bottom) data are considered. The first columns show the constraints on the $\Lambda$CDM+r model, while the next two show the results for the minimally and non-minimally coupled natural inflation, respectively. The $^\star$ indicates a derived parameter, while the constraints on $r$ are shown at 95\% confidence level.}
        \label{tab:1}
	\end{table*}

\begin{figure}
\centering
\includegraphics[width=0.49\textwidth]{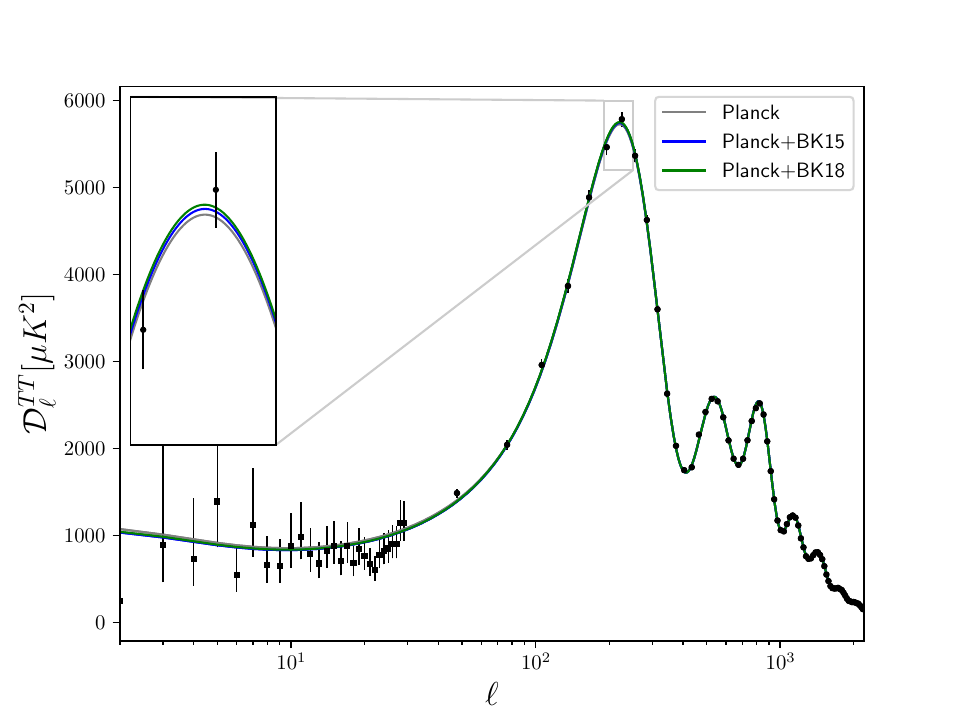}
\includegraphics[width=0.49\textwidth]{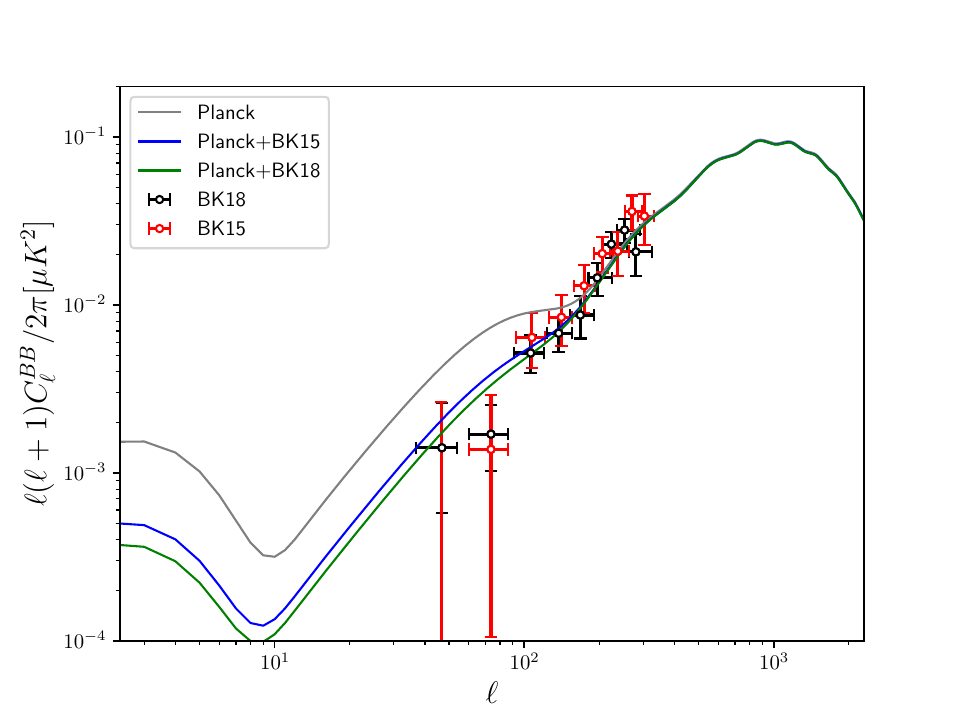}
\caption{The panel on the left-hand side shows the temperature power spectra for the best-fit parameters obtained for the non-minimally coupled natural inflation analyses, compared with the Planck estimates. In particular, the curve from the Planck+lensing results is in gray, while the Planck+BK15 is in blue, and the Planck+BK18 result is in green. We show the B-mode polarization spectra from CMB in the panel on the right-hand side for the same model. For reference, we also show the curve from Planck+lensing, while the corresponding measurements from BK15 and BK18 are shown in red and black, respectively.}\label{fig:5}
\end{figure}

A nonzero coupling leads to significant changes in the results. Table \ref{tab:1} and figs. \ref{fig:4} and \ref{fig:5} show the results for the non-minimally coupled model using the same data sets discussed above. Considering only Planck+lensing data, we note a slight preference for a negative coupling, in agreement with the model's prediction in the $n_s-r$ plane. We find $\xi=-0.00036^{+0.00070}_{-0.00014}$ and $\text{log}_{10}f/M_p=0.974^{+0.097}_{-0.28}$ at $68\%$ (C.L.), which confirms the ability of the model in alleviating the magnitude of the value for $f/M_p$, as seen in the correlation between the two parameters in fig. \ref{fig:4}, since we already obtain an upper limit on the parameter. In the same plot, we also show the impact of the BK15 and BK18 data in the evolution of this correlation. As the B-mode data are included, the contours follow a characteristic path due to the restriction imposed on $f$ for a given negative value of $\xi$. This has two important implications: first, a strong preference for a negative $\xi$ arises; for Planck+BK18 data, we are able to estimate $\xi=-0.0081^{+0.0065}_{-0.0051}$ ($68\%$ C.L.), resulting in a preference for the non-minimally coupled model. Also, the inclusion of more recent polarization data leads to a weaker constraint on the strength of the coupling, while the estimates on $f/M_p$ become tighter. Looking at the confidence contours for the three data sets in fig. \ref{fig:4}, it is possible to notice that as $\xi$ becomes more negative, the better the constraint on $\text{log}_{10}f/M_p$ will be, but with a worse estimate on $\xi$.

Table \ref{tab:1} also shows the results of the DIC criterion for all the models and data considered. We initially find that the minimally and non-minimally coupled models become slightly preferred by data compared to the $\Lambda$CDM one, but still at the level of statistical compatibility. The results change significantly when BICEP/Keck Array data is included. For the Planck+BK15 combination, we find $\Delta$DIC$=3.951$ and $\Delta$DIC$=-0.184$ for the minimally and non-minimally coupled models, respectively. These $\Delta$DIC values reflect the results of the $n_s-r$ plane shown in Fig. \ref{fig:1}, where the predictions of the minimally-coupled model only cover the $2\sigma$ region, while a small, nonzero coupling is enough for one to obtain $n_s=0.9644$, $r=0.0356$. As expected, the Planck+BK18 joint analysis excludes the minimally coupled model, as we obtain $\Delta$DIC$=10.662$, showing a strong preference for the $\Lambda$CDM model. Conversely, we obtain $\Delta$DIC$=2.869$ for the non-minimally coupled scenario, implying that the model is indeed competitive with respect to the standard model. 

The constraints obtained on the amplitude of the scalar perturbations $A_s$, the scale of symmetry breaking $f$, and the non-minimal coupling magnitude $\xi$, enable one to estimate the energy scale where the non-perturbative effects take place in the fundamental field theory, $\Lambda$, through the correlation described in eq. (\ref{eq:2.4}). Considering the $68\%$ confidence limits obtained from the Planck+BK18 data set, one obtains $\Lambda \simeq 
(1.44\, - \, 1.18) \times 10^{16}$ GeV and $\Lambda \simeq (0.46-1.16) \times 10^{16}$ GeV for minimally and non-minimally coupled models, respectively. From the phenomenological perspective, this is specially interesting since it allows to compute the predictions for the mass of the ALP in each model. Such mass is evaluated, as usual, from the second-order derivative of the potential around its vacuum expectation value (vev). In the minimally-coupled scenario, the inflationary potential assumes its simplest periodic form (\ref{eq:2.1}), which enables $m_\phi = \Lambda^2/f \simeq (2.19-2.53)\times 10^{12}$ GeV. For the non-minimally coupled model, however, the non-minimal coupling to the Ricci scalar alters the vev of the theory in a non-trivial way, $V_E(\phi)\equiv V_J(\phi)/\Omega^4(\phi)$. The numerical solution for the second-order derivative of the inflationary potential results in $m_\chi \simeq (2  - 7.68)\times 10^{13}$ GeV. Such massive ALP warrants these inflationary scenarios to escape from any stellar astrophysics limits, including those coming from star cooling and neutrino emission \cite{Marsh:2015xka}.

\section{Conclusions}\label{sec:5}

The natural inflation model is still one of the best theoretically motivated models, even decades after its conception. Although it is also a simple construction from the phenomenological point of view, the increase in the robustness of CMB data has made its viability increasingly more complex, and the recent BICEP/Keck Array CMB B-mode polarization data has excluded the model as a possible construction for inflation. Even when well-known, other inflationary models are still compatible with current data, the ‘naturalness’ and simplicity of the model are still attractive features that motivate further studies on its observational viability. 

While some extensions of the original model have been proposed~\cite{Antoniadis:2018yfq,AlHallak:2022haa,AlHallak:2022gbv,Montefalcone:2022jfw}, in this work, we focused on analyzing the consequences of the simplest possible coupling of a phenomenological axion-like field and gravity, i.e., through a quadratic term of the field non-minimally coupled to the Ricci scalar. This possibility has important implications in the slow-roll description of the model, such that a complete analysis with the current CMB cosmological data is important in order to tell if the addition of this extra theoretical ingredient is the right path to not only restore concordance with data, but also to make this model a competitive one. For the minimally coupled model, we showed that the constraints on $f/M_p$ improve when we consider additional data sets. Specifically, for the `Planck+BK15/BK18' combination, we obtained constraints on $\text{log}_{10}f/M_p$, that mitigate the superplanckian value problem but does not restore concordance with the $n_s-r$ constraints. Considering only Planck+lensing data, a preference for a small, negative coupling emerges, aligning with the model's concordance in the $n_s-r$ plane. While our analysis found $\xi=-0.00036^{+0.00070}_{-0.00014}$ and $\text{log}_{10}f/M_p=0.974^{+0.097}_{-0.28}$ at 68\% (C.L.), this preference for a non-minimal coupling becomes even stronger when we consider the BK18 data, leading to an estimate of $\xi=-0.0081^{+0.0065}_{-0.0051}$, also at $68\%$ (C.L.).

Our model comparison analysis shows that minimally and non-minimally coupled models are statistically equivalent to the $\Lambda$CDM one without the BICEP/Keck Array data. However, when considering B-mode polarization data, we find that the minimally coupled model becomes ruled out ($\Delta$DIC$=10.662$), while the non-minimally coupled model is still competitive ($\Delta$DIC$=2.869$). 

Finally, it is worth mentioning that to complement our investigation on the non-minimally coupled natural inflation scenario, one should study the impact on the reheating period when couplings of the field to other matter fields are considered and whether the presence of a non-zero $\xi$ significantly affects this era. Such analysis is in progress and will be presented in a forthcoming communication.

\section*{Acknowledgements}

FBMS and JGR acknowledge financial support from the Programa de Capacita\c{c}\~ao Institucional do Observat\'orio Nacional (PCI/ON/MCTI). GR and RS are supported by the Coordena\c{c}\~ao de Aperfei\c{c}oamento de Pessoal de N\'ivel Superior (CAPES). JSA is supported by CNPq grant No. 307683/2022-2 and Funda\c{c}\~ao de Amparo \`a Pesquisa do Estado do Rio de Janeiro (FAPERJ) grant No. 259610 (2021). We also acknowledge the use of CLASS and MontePython codes. This work was developed thanks to the use of the National Observatory Data Center (CPDON).

\bibliography{references}

\begin{thebibliography}{59}%
\makeatletter
\providecommand \@ifxundefined [1]{%
 \@ifx{#1\undefined}
}%
\providecommand \@ifnum [1]{%
 \ifnum #1\expandafter \@firstoftwo
 \else \expandafter \@secondoftwo
 \fi
}%
\providecommand \@ifx [1]{%
 \ifx #1\expandafter \@firstoftwo
 \else \expandafter \@secondoftwo
 \fi
}%
\providecommand \natexlab [1]{#1}%
\providecommand \enquote  [1]{``#1''}%
\providecommand \bibnamefont  [1]{#1}%
\providecommand \bibfnamefont [1]{#1}%
\providecommand \citenamefont [1]{#1}%
\providecommand \href@noop [0]{\@secondoftwo}%
\providecommand \href [0]{\begingroup \@sanitize@url \@href}%
\providecommand \@href[1]{\@@startlink{#1}\@@href}%
\providecommand \@@href[1]{\endgroup#1\@@endlink}%
\providecommand \@sanitize@url [0]{\catcode `\\12\catcode `\$12\catcode
  `\&12\catcode `\#12\catcode `\^12\catcode `\_12\catcode `\%12\relax}%
\providecommand \@@startlink[1]{}%
\providecommand \@@endlink[0]{}%
\providecommand \url  [0]{\begingroup\@sanitize@url \@url }%
\providecommand \@url [1]{\endgroup\@href {#1}{\urlprefix }}%
\providecommand \urlprefix  [0]{URL }%
\providecommand \Eprint [0]{\href }%
\providecommand \doibase [0]{http://dx.doi.org/}%
\providecommand \selectlanguage [0]{\@gobble}%
\providecommand \bibinfo  [0]{\@secondoftwo}%
\providecommand \bibfield  [0]{\@secondoftwo}%
\providecommand \translation [1]{[#1]}%
\providecommand \BibitemOpen [0]{}%
\providecommand \bibitemStop [0]{}%
\providecommand \bibitemNoStop [0]{.\EOS\space}%
\providecommand \EOS [0]{\spacefactor3000\relax}%
\providecommand \BibitemShut  [1]{\csname bibitem#1\endcsname}%
\let\auto@bib@innerbib\@empty
\bibitem [{\citenamefont {Starobinsky}(1980)}]{Starobinsky:1980te}%
  \BibitemOpen
  \bibfield  {author} {\bibinfo {author} {\bibfnamefont {A.~A.}\ \bibnamefont
  {Starobinsky}},\ }\href {\doibase 10.1016/0370-2693(80)90670-X} {\bibfield
  {journal} {\bibinfo  {journal} {Phys. Lett. B}\ }\textbf {\bibinfo {volume}
  {91}},\ \bibinfo {pages} {99} (\bibinfo {year} {1980})}\BibitemShut {NoStop}%
\bibitem [{\citenamefont {Guth}(1981)}]{Guth:1980zm}%
  \BibitemOpen
  \bibfield  {author} {\bibinfo {author} {\bibfnamefont {A.~H.}\ \bibnamefont
  {Guth}},\ }\href {\doibase 10.1103/PhysRevD.23.347} {\bibfield  {journal}
  {\bibinfo  {journal} {Phys. Rev. D}\ }\textbf {\bibinfo {volume} {23}},\
  \bibinfo {pages} {347} (\bibinfo {year} {1981})}\BibitemShut {NoStop}%
\bibitem [{\citenamefont {Linde}(1982)}]{Linde:1981mu}%
  \BibitemOpen
  \bibfield  {author} {\bibinfo {author} {\bibfnamefont {A.~D.}\ \bibnamefont
  {Linde}},\ }\href {\doibase 10.1016/0370-2693(82)91219-9} {\bibfield
  {journal} {\bibinfo  {journal} {Phys. Lett. B}\ }\textbf {\bibinfo {volume}
  {108}},\ \bibinfo {pages} {389} (\bibinfo {year} {1982})}\BibitemShut
  {NoStop}%
\bibitem [{\citenamefont {Albrecht}\ and\ \citenamefont
  {Steinhardt}(1982)}]{Albrecht:1982wi}%
  \BibitemOpen
  \bibfield  {author} {\bibinfo {author} {\bibfnamefont {A.}~\bibnamefont
  {Albrecht}}\ and\ \bibinfo {author} {\bibfnamefont {P.~J.}\ \bibnamefont
  {Steinhardt}},\ }\href {\doibase 10.1103/PhysRevLett.48.1220} {\bibfield
  {journal} {\bibinfo  {journal} {Phys. Rev. Lett.}\ }\textbf {\bibinfo
  {volume} {48}},\ \bibinfo {pages} {1220} (\bibinfo {year}
  {1982})}\BibitemShut {NoStop}%
\bibitem [{\citenamefont {Akrami}\ \emph {et~al.}(2020)\citenamefont {Akrami}
  \emph {et~al.}}]{Planck:2018jri}%
  \BibitemOpen
  \bibfield  {author} {\bibinfo {author} {\bibfnamefont {Y.}~\bibnamefont
  {Akrami}} \emph {et~al.} (\bibinfo {collaboration} {Planck}),\ }\href
  {\doibase 10.1051/0004-6361/201833887} {\bibfield  {journal} {\bibinfo
  {journal} {Astron. Astrophys.}\ }\textbf {\bibinfo {volume} {641}},\ \bibinfo
  {pages} {A10} (\bibinfo {year} {2020})},\ \Eprint
  {http://arxiv.org/abs/1807.06211} {arXiv:1807.06211 [astro-ph.CO]}
  \BibitemShut {NoStop}%
\bibitem [{\citenamefont {Aghanim}\ \emph
  {et~al.}(2020{\natexlab{a}})\citenamefont {Aghanim} \emph
  {et~al.}}]{Planck:2018nkj}%
  \BibitemOpen
  \bibfield  {author} {\bibinfo {author} {\bibfnamefont {N.}~\bibnamefont
  {Aghanim}} \emph {et~al.} (\bibinfo {collaboration} {Planck}),\ }\href
  {\doibase 10.1051/0004-6361/201833880} {\bibfield  {journal} {\bibinfo
  {journal} {Astron. Astrophys.}\ }\textbf {\bibinfo {volume} {641}},\ \bibinfo
  {pages} {A1} (\bibinfo {year} {2020}{\natexlab{a}})},\ \Eprint
  {http://arxiv.org/abs/1807.06205} {arXiv:1807.06205 [astro-ph.CO]}
  \BibitemShut {NoStop}%
\bibitem [{\citenamefont {Aghanim}\ \emph
  {et~al.}(2020{\natexlab{b}})\citenamefont {Aghanim} \emph
  {et~al.}}]{Planck:2018vyg}%
  \BibitemOpen
  \bibfield  {author} {\bibinfo {author} {\bibfnamefont {N.}~\bibnamefont
  {Aghanim}} \emph {et~al.} (\bibinfo {collaboration} {Planck}),\ }\href
  {\doibase 10.1051/0004-6361/201833910} {\bibfield  {journal} {\bibinfo
  {journal} {Astron. Astrophys.}\ }\textbf {\bibinfo {volume} {641}},\ \bibinfo
  {pages} {A6} (\bibinfo {year} {2020}{\natexlab{b}})},\ \bibinfo {note}
  {[Erratum: Astron.Astrophys. 652, C4 (2021)]},\ \Eprint
  {http://arxiv.org/abs/1807.06209} {arXiv:1807.06209 [astro-ph.CO]}
  \BibitemShut {NoStop}%
\bibitem [{\citenamefont {Futamase}\ and\ \citenamefont
  {Maeda}(1989)}]{Futamase:1987ua}%
  \BibitemOpen
  \bibfield  {author} {\bibinfo {author} {\bibfnamefont {T.}~\bibnamefont
  {Futamase}}\ and\ \bibinfo {author} {\bibfnamefont {K.-i.}\ \bibnamefont
  {Maeda}},\ }\href {\doibase 10.1103/PhysRevD.39.399} {\bibfield  {journal}
  {\bibinfo  {journal} {Phys. Rev. D}\ }\textbf {\bibinfo {volume} {39}},\
  \bibinfo {pages} {399} (\bibinfo {year} {1989})}\BibitemShut {NoStop}%
\bibitem [{\citenamefont {Fakir}\ and\ \citenamefont
  {Unruh}(1990)}]{Fakir:1990eg}%
  \BibitemOpen
  \bibfield  {author} {\bibinfo {author} {\bibfnamefont {R.}~\bibnamefont
  {Fakir}}\ and\ \bibinfo {author} {\bibfnamefont {W.~G.}\ \bibnamefont
  {Unruh}},\ }\href {\doibase 10.1103/PhysRevD.41.1783} {\bibfield  {journal}
  {\bibinfo  {journal} {Phys. Rev. D}\ }\textbf {\bibinfo {volume} {41}},\
  \bibinfo {pages} {1783} (\bibinfo {year} {1990})}\BibitemShut {NoStop}%
\bibitem [{\citenamefont {Faraoni}(1996)}]{Faraoni:1996rf}%
  \BibitemOpen
  \bibfield  {author} {\bibinfo {author} {\bibfnamefont {V.}~\bibnamefont
  {Faraoni}},\ }\href {\doibase 10.1103/PhysRevD.53.6813} {\bibfield  {journal}
  {\bibinfo  {journal} {Phys. Rev. D}\ }\textbf {\bibinfo {volume} {53}},\
  \bibinfo {pages} {6813} (\bibinfo {year} {1996})},\ \Eprint
  {http://arxiv.org/abs/astro-ph/9602111} {arXiv:astro-ph/9602111} \BibitemShut
  {NoStop}%
\bibitem [{\citenamefont {Tsujikawa}\ and\ \citenamefont
  {Gumjudpai}(2004)}]{Tsujikawa:2004my}%
  \BibitemOpen
  \bibfield  {author} {\bibinfo {author} {\bibfnamefont {S.}~\bibnamefont
  {Tsujikawa}}\ and\ \bibinfo {author} {\bibfnamefont {B.}~\bibnamefont
  {Gumjudpai}},\ }\href {\doibase 10.1103/PhysRevD.69.123523} {\bibfield
  {journal} {\bibinfo  {journal} {Phys. Rev. D}\ }\textbf {\bibinfo {volume}
  {69}},\ \bibinfo {pages} {123523} (\bibinfo {year} {2004})},\ \Eprint
  {http://arxiv.org/abs/astro-ph/0402185} {arXiv:astro-ph/0402185} \BibitemShut
  {NoStop}%
\bibitem [{\citenamefont {Nozari}\ and\ \citenamefont
  {Sadatian}(2008)}]{Nozari:2007eq}%
  \BibitemOpen
  \bibfield  {author} {\bibinfo {author} {\bibfnamefont {K.}~\bibnamefont
  {Nozari}}\ and\ \bibinfo {author} {\bibfnamefont {S.~D.}\ \bibnamefont
  {Sadatian}},\ }\href {\doibase 10.1142/S0217732308026698} {\bibfield
  {journal} {\bibinfo  {journal} {Mod. Phys. Lett. A}\ }\textbf {\bibinfo
  {volume} {23}},\ \bibinfo {pages} {2933} (\bibinfo {year} {2008})},\ \Eprint
  {http://arxiv.org/abs/0710.0058} {arXiv:0710.0058 [astro-ph]} \BibitemShut
  {NoStop}%
\bibitem [{\citenamefont {Bauer}\ and\ \citenamefont
  {Demir}(2008)}]{Bauer:2008zj}%
  \BibitemOpen
  \bibfield  {author} {\bibinfo {author} {\bibfnamefont {F.}~\bibnamefont
  {Bauer}}\ and\ \bibinfo {author} {\bibfnamefont {D.~A.}\ \bibnamefont
  {Demir}},\ }\href {\doibase 10.1016/j.physletb.2008.06.014} {\bibfield
  {journal} {\bibinfo  {journal} {Phys. Lett. B}\ }\textbf {\bibinfo {volume}
  {665}},\ \bibinfo {pages} {222} (\bibinfo {year} {2008})},\ \Eprint
  {http://arxiv.org/abs/0803.2664} {arXiv:0803.2664 [hep-ph]} \BibitemShut
  {NoStop}%
\bibitem [{\citenamefont {Okada}\ \emph {et~al.}(2010)\citenamefont {Okada},
  \citenamefont {Rehman},\ and\ \citenamefont {Shafi}}]{Okada:2010jf}%
  \BibitemOpen
  \bibfield  {author} {\bibinfo {author} {\bibfnamefont {N.}~\bibnamefont
  {Okada}}, \bibinfo {author} {\bibfnamefont {M.~U.}\ \bibnamefont {Rehman}}, \
  and\ \bibinfo {author} {\bibfnamefont {Q.}~\bibnamefont {Shafi}},\ }\href
  {\doibase 10.1103/PhysRevD.82.043502} {\bibfield  {journal} {\bibinfo
  {journal} {Phys. Rev. D}\ }\textbf {\bibinfo {volume} {82}},\ \bibinfo
  {pages} {043502} (\bibinfo {year} {2010})},\ \Eprint
  {http://arxiv.org/abs/1005.5161} {arXiv:1005.5161 [hep-ph]} \BibitemShut
  {NoStop}%
\bibitem [{\citenamefont {Hertzberg}(2010)}]{Hertzberg:2010dc}%
  \BibitemOpen
  \bibfield  {author} {\bibinfo {author} {\bibfnamefont {M.~P.}\ \bibnamefont
  {Hertzberg}},\ }\href {\doibase 10.1007/JHEP11(2010)023} {\bibfield
  {journal} {\bibinfo  {journal} {JHEP}\ }\textbf {\bibinfo {volume} {11}},\
  \bibinfo {pages} {023} (\bibinfo {year} {2010})},\ \Eprint
  {http://arxiv.org/abs/1002.2995} {arXiv:1002.2995 [hep-ph]} \BibitemShut
  {NoStop}%
\bibitem [{\citenamefont {Tenkanen}(2017)}]{Tenkanen:2017jih}%
  \BibitemOpen
  \bibfield  {author} {\bibinfo {author} {\bibfnamefont {T.}~\bibnamefont
  {Tenkanen}},\ }\href {\doibase 10.1088/1475-7516/2017/12/001} {\bibfield
  {journal} {\bibinfo  {journal} {JCAP}\ }\textbf {\bibinfo {volume} {12}},\
  \bibinfo {pages} {001} (\bibinfo {year} {2017})},\ \Eprint
  {http://arxiv.org/abs/1710.02758} {arXiv:1710.02758 [astro-ph.CO]}
  \BibitemShut {NoStop}%
\bibitem [{\citenamefont {Rodrigues}\ \emph {et~al.}(2020)\citenamefont
  {Rodrigues}, \citenamefont {Benetti}, \citenamefont {Campista},\ and\
  \citenamefont {Alcaniz}}]{Rodrigues:2020dod}%
  \BibitemOpen
  \bibfield  {author} {\bibinfo {author} {\bibfnamefont {J.~G.}\ \bibnamefont
  {Rodrigues}}, \bibinfo {author} {\bibfnamefont {M.}~\bibnamefont {Benetti}},
  \bibinfo {author} {\bibfnamefont {M.}~\bibnamefont {Campista}}, \ and\
  \bibinfo {author} {\bibfnamefont {J.}~\bibnamefont {Alcaniz}},\ }\href
  {\doibase 10.1088/1475-7516/2020/07/007} {\bibfield  {journal} {\bibinfo
  {journal} {JCAP}\ }\textbf {\bibinfo {volume} {07}},\ \bibinfo {pages} {007}
  (\bibinfo {year} {2020})},\ \Eprint {http://arxiv.org/abs/2002.05154}
  {arXiv:2002.05154 [astro-ph.CO]} \BibitemShut {NoStop}%
\bibitem [{\citenamefont {Rodrigues}\ \emph {et~al.}(2021)\citenamefont
  {Rodrigues}, \citenamefont {Benetti},\ and\ \citenamefont
  {Alcaniz}}]{Rodrigues:2021txa}%
  \BibitemOpen
  \bibfield  {author} {\bibinfo {author} {\bibfnamefont {J.~G.}\ \bibnamefont
  {Rodrigues}}, \bibinfo {author} {\bibfnamefont {M.}~\bibnamefont {Benetti}},
  \ and\ \bibinfo {author} {\bibfnamefont {J.~S.}\ \bibnamefont {Alcaniz}},\
  }\href {\doibase 10.1007/JHEP11(2021)091} {\bibfield  {journal} {\bibinfo
  {journal} {JHEP}\ }\textbf {\bibinfo {volume} {11}},\ \bibinfo {pages} {091}
  (\bibinfo {year} {2021})},\ \Eprint {http://arxiv.org/abs/2105.07009}
  {arXiv:2105.07009 [hep-ph]} \BibitemShut {NoStop}%
\bibitem [{\citenamefont {Campista}\ \emph {et~al.}(2017)\citenamefont
  {Campista}, \citenamefont {Benetti},\ and\ \citenamefont
  {Alcaniz}}]{Campista:2017ovq}%
  \BibitemOpen
  \bibfield  {author} {\bibinfo {author} {\bibfnamefont {M.}~\bibnamefont
  {Campista}}, \bibinfo {author} {\bibfnamefont {M.}~\bibnamefont {Benetti}}, \
  and\ \bibinfo {author} {\bibfnamefont {J.}~\bibnamefont {Alcaniz}},\ }\href
  {\doibase 10.1088/1475-7516/2017/09/010} {\bibfield  {journal} {\bibinfo
  {journal} {JCAP}\ }\textbf {\bibinfo {volume} {09}},\ \bibinfo {pages} {010}
  (\bibinfo {year} {2017})},\ \Eprint {http://arxiv.org/abs/1705.08877}
  {arXiv:1705.08877 [astro-ph.CO]} \BibitemShut {NoStop}%
\bibitem [{\citenamefont {dos Santos}\ \emph {et~al.}(2022)\citenamefont {dos
  Santos}, \citenamefont {Santos~da Costa}, \citenamefont {Silva},
  \citenamefont {Benetti},\ and\ \citenamefont {Alcaniz}}]{dosSantos:2021vis}%
  \BibitemOpen
  \bibfield  {author} {\bibinfo {author} {\bibfnamefont {F.~B.~M.}\
  \bibnamefont {dos Santos}}, \bibinfo {author} {\bibfnamefont
  {S.}~\bibnamefont {Santos~da Costa}}, \bibinfo {author} {\bibfnamefont
  {R.}~\bibnamefont {Silva}}, \bibinfo {author} {\bibfnamefont
  {M.}~\bibnamefont {Benetti}}, \ and\ \bibinfo {author} {\bibfnamefont
  {J.}~\bibnamefont {Alcaniz}},\ }\href {\doibase
  10.1088/1475-7516/2022/06/001} {\bibfield  {journal} {\bibinfo  {journal}
  {JCAP}\ }\textbf {\bibinfo {volume} {06}},\ \bibinfo {pages} {001} (\bibinfo
  {year} {2022})},\ \Eprint {http://arxiv.org/abs/2110.14758} {arXiv:2110.14758
  [astro-ph.CO]} \BibitemShut {NoStop}%
\bibitem [{\citenamefont {Santos}\ \emph {et~al.}(2023)\citenamefont {Santos},
  \citenamefont {Silva},\ and\ \citenamefont {Alcaniz}}]{Santos:2023hhk}%
  \BibitemOpen
  \bibfield  {author} {\bibinfo {author} {\bibfnamefont {F.~B. M.~d.}\
  \bibnamefont {Santos}}, \bibinfo {author} {\bibfnamefont {R.}~\bibnamefont
  {Silva}}, \ and\ \bibinfo {author} {\bibfnamefont {J.~S.}\ \bibnamefont
  {Alcaniz}},\ }\href {\doibase 10.1088/1475-7516/2023/07/027} {\bibfield
  {journal} {\bibinfo  {journal} {JCAP}\ }\textbf {\bibinfo {volume} {07}},\
  \bibinfo {pages} {027} (\bibinfo {year} {2023})},\ \Eprint
  {http://arxiv.org/abs/2306.07260} {arXiv:2306.07260 [astro-ph.CO]}
  \BibitemShut {NoStop}%
\bibitem [{\citenamefont {Bezrukov}\ and\ \citenamefont
  {Shaposhnikov}(2008)}]{Bezrukov:2007ep}%
  \BibitemOpen
  \bibfield  {author} {\bibinfo {author} {\bibfnamefont {F.~L.}\ \bibnamefont
  {Bezrukov}}\ and\ \bibinfo {author} {\bibfnamefont {M.}~\bibnamefont
  {Shaposhnikov}},\ }\href {\doibase 10.1016/j.physletb.2007.11.072} {\bibfield
   {journal} {\bibinfo  {journal} {Phys. Lett. B}\ }\textbf {\bibinfo {volume}
  {659}},\ \bibinfo {pages} {703} (\bibinfo {year} {2008})},\ \Eprint
  {http://arxiv.org/abs/0710.3755} {arXiv:0710.3755 [hep-th]} \BibitemShut
  {NoStop}%
\bibitem [{\citenamefont {Freese}\ \emph {et~al.}(1990)\citenamefont {Freese},
  \citenamefont {Frieman},\ and\ \citenamefont {Olinto}}]{Freese:1990rb}%
  \BibitemOpen
  \bibfield  {author} {\bibinfo {author} {\bibfnamefont {K.}~\bibnamefont
  {Freese}}, \bibinfo {author} {\bibfnamefont {J.~A.}\ \bibnamefont {Frieman}},
  \ and\ \bibinfo {author} {\bibfnamefont {A.~V.}\ \bibnamefont {Olinto}},\
  }\href {\doibase 10.1103/PhysRevLett.65.3233} {\bibfield  {journal} {\bibinfo
   {journal} {Phys. Rev. Lett.}\ }\textbf {\bibinfo {volume} {65}},\ \bibinfo
  {pages} {3233} (\bibinfo {year} {1990})}\BibitemShut {NoStop}%
\bibitem [{\citenamefont {Barkats}\ \emph {et~al.}(2014)\citenamefont {Barkats}
  \emph {et~al.}}]{BICEP1:2013sbv}%
  \BibitemOpen
  \bibfield  {author} {\bibinfo {author} {\bibfnamefont {D.}~\bibnamefont
  {Barkats}} \emph {et~al.} (\bibinfo {collaboration} {BICEP1}),\ }\href
  {\doibase 10.1088/0004-637X/783/2/67} {\bibfield  {journal} {\bibinfo
  {journal} {Astrophys. J.}\ }\textbf {\bibinfo {volume} {783}},\ \bibinfo
  {pages} {67} (\bibinfo {year} {2014})},\ \Eprint
  {http://arxiv.org/abs/1310.1422} {arXiv:1310.1422 [astro-ph.CO]} \BibitemShut
  {NoStop}%
\bibitem [{\citenamefont {Ade}\ \emph {et~al.}(2018)\citenamefont {Ade} \emph
  {et~al.}}]{BICEP2:2018kqh}%
  \BibitemOpen
  \bibfield  {author} {\bibinfo {author} {\bibfnamefont {P.~A.~R.}\
  \bibnamefont {Ade}} \emph {et~al.} (\bibinfo {collaboration} {BICEP2, Keck
  Array}),\ }\href {\doibase 10.1103/PhysRevLett.121.221301} {\bibfield
  {journal} {\bibinfo  {journal} {Phys. Rev. Lett.}\ }\textbf {\bibinfo
  {volume} {121}},\ \bibinfo {pages} {221301} (\bibinfo {year} {2018})},\
  \Eprint {http://arxiv.org/abs/1810.05216} {arXiv:1810.05216 [astro-ph.CO]}
  \BibitemShut {NoStop}%
\bibitem [{\citenamefont {Ade}\ \emph {et~al.}(2021)\citenamefont {Ade} \emph
  {et~al.}}]{BICEP:2021xfz}%
  \BibitemOpen
  \bibfield  {author} {\bibinfo {author} {\bibfnamefont {P.~A.~R.}\
  \bibnamefont {Ade}} \emph {et~al.} (\bibinfo {collaboration} {BICEP, Keck}),\
  }\href {\doibase 10.1103/PhysRevLett.127.151301} {\bibfield  {journal}
  {\bibinfo  {journal} {Phys. Rev. Lett.}\ }\textbf {\bibinfo {volume} {127}},\
  \bibinfo {pages} {151301} (\bibinfo {year} {2021})},\ \Eprint
  {http://arxiv.org/abs/2110.00483} {arXiv:2110.00483 [astro-ph.CO]}
  \BibitemShut {NoStop}%
\bibitem [{\citenamefont {Gerbino}\ \emph {et~al.}(2017)\citenamefont
  {Gerbino}, \citenamefont {Freese}, \citenamefont {Vagnozzi}, \citenamefont
  {Lattanzi}, \citenamefont {Mena}, \citenamefont {Giusarma},\ and\
  \citenamefont {Ho}}]{Gerbino:2016sgw}%
  \BibitemOpen
  \bibfield  {author} {\bibinfo {author} {\bibfnamefont {M.}~\bibnamefont
  {Gerbino}}, \bibinfo {author} {\bibfnamefont {K.}~\bibnamefont {Freese}},
  \bibinfo {author} {\bibfnamefont {S.}~\bibnamefont {Vagnozzi}}, \bibinfo
  {author} {\bibfnamefont {M.}~\bibnamefont {Lattanzi}}, \bibinfo {author}
  {\bibfnamefont {O.}~\bibnamefont {Mena}}, \bibinfo {author} {\bibfnamefont
  {E.}~\bibnamefont {Giusarma}}, \ and\ \bibinfo {author} {\bibfnamefont
  {S.}~\bibnamefont {Ho}},\ }\href {\doibase 10.1103/PhysRevD.95.043512}
  {\bibfield  {journal} {\bibinfo  {journal} {Phys. Rev. D}\ }\textbf {\bibinfo
  {volume} {95}},\ \bibinfo {pages} {043512} (\bibinfo {year} {2017})},\
  \Eprint {http://arxiv.org/abs/1610.08830} {arXiv:1610.08830 [astro-ph.CO]}
  \BibitemShut {NoStop}%
\bibitem [{\citenamefont {Ferreira}\ \emph {et~al.}(2018)\citenamefont
  {Ferreira}, \citenamefont {Notari},\ and\ \citenamefont
  {Simeon}}]{Ferreira:2018nav}%
  \BibitemOpen
  \bibfield  {author} {\bibinfo {author} {\bibfnamefont {R.~Z.}\ \bibnamefont
  {Ferreira}}, \bibinfo {author} {\bibfnamefont {A.}~\bibnamefont {Notari}}, \
  and\ \bibinfo {author} {\bibfnamefont {G.}~\bibnamefont {Simeon}},\ }\href
  {\doibase 10.1088/1475-7516/2018/11/021} {\bibfield  {journal} {\bibinfo
  {journal} {JCAP}\ }\textbf {\bibinfo {volume} {11}},\ \bibinfo {pages} {021}
  (\bibinfo {year} {2018})},\ \Eprint {http://arxiv.org/abs/1806.05511}
  {arXiv:1806.05511 [astro-ph.CO]} \BibitemShut {NoStop}%
\bibitem [{\citenamefont {Reyimuaji}\ and\ \citenamefont
  {Zhang}(2021)}]{Reyimuaji:2020goi}%
  \BibitemOpen
  \bibfield  {author} {\bibinfo {author} {\bibfnamefont {Y.}~\bibnamefont
  {Reyimuaji}}\ and\ \bibinfo {author} {\bibfnamefont {X.}~\bibnamefont
  {Zhang}},\ }\href {\doibase 10.1088/1475-7516/2021/03/059} {\bibfield
  {journal} {\bibinfo  {journal} {JCAP}\ }\textbf {\bibinfo {volume} {03}},\
  \bibinfo {pages} {059} (\bibinfo {year} {2021})},\ \Eprint
  {http://arxiv.org/abs/2012.14248} {arXiv:2012.14248 [astro-ph.CO]}
  \BibitemShut {NoStop}%
\bibitem [{\citenamefont {Bostan}(2023)}]{Bostan:2022swq}%
  \BibitemOpen
  \bibfield  {author} {\bibinfo {author} {\bibfnamefont {N.}~\bibnamefont
  {Bostan}},\ }\href {\doibase 10.1088/1475-7516/2023/02/063} {\bibfield
  {journal} {\bibinfo  {journal} {JCAP}\ }\textbf {\bibinfo {volume} {02}},\
  \bibinfo {pages} {063} (\bibinfo {year} {2023})},\ \Eprint
  {http://arxiv.org/abs/2209.02434} {arXiv:2209.02434 [astro-ph.CO]}
  \BibitemShut {NoStop}%
\bibitem [{\citenamefont {Bostan}\ and\ \citenamefont
  {Roy~Choudhury}(2023)}]{Bostan:2023ped}%
  \BibitemOpen
  \bibfield  {author} {\bibinfo {author} {\bibfnamefont {N.}~\bibnamefont
  {Bostan}}\ and\ \bibinfo {author} {\bibfnamefont {S.}~\bibnamefont
  {Roy~Choudhury}},\ }\href@noop {} {\  (\bibinfo {year} {2023})},\ \Eprint
  {http://arxiv.org/abs/2310.01491} {arXiv:2310.01491 [astro-ph.CO]}
  \BibitemShut {NoStop}%
\bibitem [{\citenamefont {Kim}(1987)}]{Kim:1986ax}%
  \BibitemOpen
  \bibfield  {author} {\bibinfo {author} {\bibfnamefont {J.~E.}\ \bibnamefont
  {Kim}},\ }\href {\doibase 10.1016/0370-1573(87)90017-2} {\bibfield  {journal}
  {\bibinfo  {journal} {Phys. Rept.}\ }\textbf {\bibinfo {volume} {150}},\
  \bibinfo {pages} {1} (\bibinfo {year} {1987})}\BibitemShut {NoStop}%
\bibitem [{\citenamefont {Svrcek}\ and\ \citenamefont
  {Witten}(2006)}]{Svrcek:2006yi}%
  \BibitemOpen
  \bibfield  {author} {\bibinfo {author} {\bibfnamefont {P.}~\bibnamefont
  {Svrcek}}\ and\ \bibinfo {author} {\bibfnamefont {E.}~\bibnamefont
  {Witten}},\ }\href {\doibase 10.1088/1126-6708/2006/06/051} {\bibfield
  {journal} {\bibinfo  {journal} {JHEP}\ }\textbf {\bibinfo {volume} {06}},\
  \bibinfo {pages} {051} (\bibinfo {year} {2006})},\ \Eprint
  {http://arxiv.org/abs/hep-th/0605206} {arXiv:hep-th/0605206} \BibitemShut
  {NoStop}%
\bibitem [{\citenamefont {Peccei}\ and\ \citenamefont
  {Quinn}(1977)}]{Peccei:1977hh}%
  \BibitemOpen
  \bibfield  {author} {\bibinfo {author} {\bibfnamefont {R.~D.}\ \bibnamefont
  {Peccei}}\ and\ \bibinfo {author} {\bibfnamefont {H.~R.}\ \bibnamefont
  {Quinn}},\ }\href {\doibase 10.1103/PhysRevLett.38.1440} {\bibfield
  {journal} {\bibinfo  {journal} {Phys. Rev. Lett.}\ }\textbf {\bibinfo
  {volume} {38}},\ \bibinfo {pages} {1440} (\bibinfo {year}
  {1977})}\BibitemShut {NoStop}%
\bibitem [{\citenamefont {Wilczek}(1978)}]{Wilczek:1977pj}%
  \BibitemOpen
  \bibfield  {author} {\bibinfo {author} {\bibfnamefont {F.}~\bibnamefont
  {Wilczek}},\ }\href {\doibase 10.1103/PhysRevLett.40.279} {\bibfield
  {journal} {\bibinfo  {journal} {Phys. Rev. Lett.}\ }\textbf {\bibinfo
  {volume} {40}},\ \bibinfo {pages} {279} (\bibinfo {year} {1978})}\BibitemShut
  {NoStop}%
\bibitem [{\citenamefont {Weinberg}(1978)}]{Weinberg:1977ma}%
  \BibitemOpen
  \bibfield  {author} {\bibinfo {author} {\bibfnamefont {S.}~\bibnamefont
  {Weinberg}},\ }\href {\doibase 10.1103/PhysRevLett.40.223} {\bibfield
  {journal} {\bibinfo  {journal} {Phys. Rev. Lett.}\ }\textbf {\bibinfo
  {volume} {40}},\ \bibinfo {pages} {223} (\bibinfo {year} {1978})}\BibitemShut
  {NoStop}%
\bibitem [{\citenamefont {Kim}(1979)}]{Kim:1979if}%
  \BibitemOpen
  \bibfield  {author} {\bibinfo {author} {\bibfnamefont {J.~E.}\ \bibnamefont
  {Kim}},\ }\href {\doibase 10.1103/PhysRevLett.43.103} {\bibfield  {journal}
  {\bibinfo  {journal} {Phys. Rev. Lett.}\ }\textbf {\bibinfo {volume} {43}},\
  \bibinfo {pages} {103} (\bibinfo {year} {1979})}\BibitemShut {NoStop}%
\bibitem [{\citenamefont {Shifman}\ \emph {et~al.}(1980)\citenamefont
  {Shifman}, \citenamefont {Vainshtein},\ and\ \citenamefont
  {Zakharov}}]{Shifman:1979if}%
  \BibitemOpen
  \bibfield  {author} {\bibinfo {author} {\bibfnamefont {M.~A.}\ \bibnamefont
  {Shifman}}, \bibinfo {author} {\bibfnamefont {A.~I.}\ \bibnamefont
  {Vainshtein}}, \ and\ \bibinfo {author} {\bibfnamefont {V.~I.}\ \bibnamefont
  {Zakharov}},\ }\href {\doibase 10.1016/0550-3213(80)90209-6} {\bibfield
  {journal} {\bibinfo  {journal} {Nucl. Phys. B}\ }\textbf {\bibinfo {volume}
  {166}},\ \bibinfo {pages} {493} (\bibinfo {year} {1980})}\BibitemShut
  {NoStop}%
\bibitem [{\citenamefont {Dine}\ \emph {et~al.}(1981)\citenamefont {Dine},
  \citenamefont {Fischler},\ and\ \citenamefont {Srednicki}}]{Dine:1981rt}%
  \BibitemOpen
  \bibfield  {author} {\bibinfo {author} {\bibfnamefont {M.}~\bibnamefont
  {Dine}}, \bibinfo {author} {\bibfnamefont {W.}~\bibnamefont {Fischler}}, \
  and\ \bibinfo {author} {\bibfnamefont {M.}~\bibnamefont {Srednicki}},\ }\href
  {\doibase 10.1016/0370-2693(81)90590-6} {\bibfield  {journal} {\bibinfo
  {journal} {Phys. Lett. B}\ }\textbf {\bibinfo {volume} {104}},\ \bibinfo
  {pages} {199} (\bibinfo {year} {1981})}\BibitemShut {NoStop}%
\bibitem [{\citenamefont {Zhitnitsky}(1980)}]{Zhitnitsky:1980tq}%
  \BibitemOpen
  \bibfield  {author} {\bibinfo {author} {\bibfnamefont {A.~R.}\ \bibnamefont
  {Zhitnitsky}},\ }\href@noop {} {\bibfield  {journal} {\bibinfo  {journal}
  {Sov. J. Nucl. Phys.}\ }\textbf {\bibinfo {volume} {31}},\ \bibinfo {pages}
  {260} (\bibinfo {year} {1980})}\BibitemShut {NoStop}%
\bibitem [{\citenamefont {Birrell}\ and\ \citenamefont
  {Davies}(1984)}]{Birrell:1982ix}%
  \BibitemOpen
  \bibfield  {author} {\bibinfo {author} {\bibfnamefont {N.~D.}\ \bibnamefont
  {Birrell}}\ and\ \bibinfo {author} {\bibfnamefont {P.~C.~W.}\ \bibnamefont
  {Davies}},\ }\href {\doibase 10.1017/CBO9780511622632} {\emph {\bibinfo
  {title} {{Quantum Fields in Curved Space}}}},\ Cambridge Monographs on
  Mathematical Physics\ (\bibinfo  {publisher} {Cambridge Univ. Press},\
  \bibinfo {address} {Cambridge, UK},\ \bibinfo {year} {1984})\BibitemShut
  {NoStop}%
\bibitem [{\citenamefont {Accioly}\ \emph {et~al.}(1993)\citenamefont
  {Accioly}, \citenamefont {Wichoski}, \citenamefont {Kwok},\ and\
  \citenamefont {Pereira~da Silva}}]{Accioly:1993kc}%
  \BibitemOpen
  \bibfield  {author} {\bibinfo {author} {\bibfnamefont {A.~J.}\ \bibnamefont
  {Accioly}}, \bibinfo {author} {\bibfnamefont {U.~F.}\ \bibnamefont
  {Wichoski}}, \bibinfo {author} {\bibfnamefont {S.~F.}\ \bibnamefont {Kwok}},
  \ and\ \bibinfo {author} {\bibfnamefont {N.~L.~P.}\ \bibnamefont {Pereira~da
  Silva}},\ }\href {\doibase 10.1088/0264-9381/10/12/001} {\bibfield  {journal}
  {\bibinfo  {journal} {Class. Quant. Grav.}\ }\textbf {\bibinfo {volume}
  {10}},\ \bibinfo {pages} {L215} (\bibinfo {year} {1993})}\BibitemShut
  {NoStop}%
\bibitem [{\citenamefont {Faraoni}\ \emph {et~al.}(1999)\citenamefont
  {Faraoni}, \citenamefont {Gunzig},\ and\ \citenamefont
  {Nardone}}]{Faraoni:1998qx}%
  \BibitemOpen
  \bibfield  {author} {\bibinfo {author} {\bibfnamefont {V.}~\bibnamefont
  {Faraoni}}, \bibinfo {author} {\bibfnamefont {E.}~\bibnamefont {Gunzig}}, \
  and\ \bibinfo {author} {\bibfnamefont {P.}~\bibnamefont {Nardone}},\
  }\href@noop {} {\bibfield  {journal} {\bibinfo  {journal} {Fund. Cosmic
  Phys.}\ }\textbf {\bibinfo {volume} {20}},\ \bibinfo {pages} {121} (\bibinfo
  {year} {1999})},\ \Eprint {http://arxiv.org/abs/gr-qc/9811047}
  {arXiv:gr-qc/9811047 [gr-qc]} \BibitemShut {NoStop}%
\bibitem [{\citenamefont {Kallosh}\ \emph {et~al.}(2014)\citenamefont
  {Kallosh}, \citenamefont {Linde},\ and\ \citenamefont
  {Roest}}]{Kallosh:2013tua}%
  \BibitemOpen
  \bibfield  {author} {\bibinfo {author} {\bibfnamefont {R.}~\bibnamefont
  {Kallosh}}, \bibinfo {author} {\bibfnamefont {A.}~\bibnamefont {Linde}}, \
  and\ \bibinfo {author} {\bibfnamefont {D.}~\bibnamefont {Roest}},\ }\href
  {\doibase 10.1103/PhysRevLett.112.011303} {\bibfield  {journal} {\bibinfo
  {journal} {Phys. Rev. Lett.}\ }\textbf {\bibinfo {volume} {112}},\ \bibinfo
  {pages} {011303} (\bibinfo {year} {2014})},\ \Eprint
  {http://arxiv.org/abs/1310.3950} {arXiv:1310.3950 [hep-th]} \BibitemShut
  {NoStop}%
\bibitem [{\citenamefont {Kallosh}\ and\ \citenamefont
  {Linde}(2013)}]{Kallosh:2013hoa}%
  \BibitemOpen
  \bibfield  {author} {\bibinfo {author} {\bibfnamefont {R.}~\bibnamefont
  {Kallosh}}\ and\ \bibinfo {author} {\bibfnamefont {A.}~\bibnamefont
  {Linde}},\ }\href {\doibase 10.1088/1475-7516/2013/07/002} {\bibfield
  {journal} {\bibinfo  {journal} {JCAP}\ }\textbf {\bibinfo {volume} {1307}},\
  \bibinfo {pages} {002} (\bibinfo {year} {2013})},\ \Eprint
  {http://arxiv.org/abs/1306.5220} {arXiv:1306.5220 [hep-th]} \BibitemShut
  {NoStop}%
\bibitem [{\citenamefont {Kallosh}\ \emph {et~al.}(2013)\citenamefont
  {Kallosh}, \citenamefont {Linde},\ and\ \citenamefont
  {Roest}}]{Kallosh:2013yoa}%
  \BibitemOpen
  \bibfield  {author} {\bibinfo {author} {\bibfnamefont {R.}~\bibnamefont
  {Kallosh}}, \bibinfo {author} {\bibfnamefont {A.}~\bibnamefont {Linde}}, \
  and\ \bibinfo {author} {\bibfnamefont {D.}~\bibnamefont {Roest}},\ }\href
  {\doibase 10.1007/JHEP11(2013)198} {\bibfield  {journal} {\bibinfo  {journal}
  {JHEP}\ }\textbf {\bibinfo {volume} {11}},\ \bibinfo {pages} {198} (\bibinfo
  {year} {2013})},\ \Eprint {http://arxiv.org/abs/1311.0472} {arXiv:1311.0472
  [hep-th]} \BibitemShut {NoStop}%
\bibitem [{\citenamefont {Giar\`e}\ \emph {et~al.}(2023)\citenamefont
  {Giar\`e}, \citenamefont {De~Angelis}, \citenamefont {van~de Bruck},\ and\
  \citenamefont {Di~Valentino}}]{Giare:2023kiv}%
  \BibitemOpen
  \bibfield  {author} {\bibinfo {author} {\bibfnamefont {W.}~\bibnamefont
  {Giar\`e}}, \bibinfo {author} {\bibfnamefont {M.}~\bibnamefont {De~Angelis}},
  \bibinfo {author} {\bibfnamefont {C.}~\bibnamefont {van~de Bruck}}, \ and\
  \bibinfo {author} {\bibfnamefont {E.}~\bibnamefont {Di~Valentino}},\ }\href
  {\doibase 10.1088/1475-7516/2023/12/014} {\bibfield  {journal} {\bibinfo
  {journal} {JCAP}\ }\textbf {\bibinfo {volume} {12}},\ \bibinfo {pages} {014}
  (\bibinfo {year} {2023})},\ \Eprint {http://arxiv.org/abs/2306.12414}
  {arXiv:2306.12414 [astro-ph.CO]} \BibitemShut {NoStop}%
\bibitem [{\citenamefont {Lesgourgues}(2011)}]{Lesgourgues:2011re}%
  \BibitemOpen
  \bibfield  {author} {\bibinfo {author} {\bibfnamefont {J.}~\bibnamefont
  {Lesgourgues}},\ }\href@noop {} {\  (\bibinfo {year} {2011})},\ \Eprint
  {http://arxiv.org/abs/1104.2932} {arXiv:1104.2932 [astro-ph.IM]} \BibitemShut
  {NoStop}%
\bibitem [{\citenamefont {Blas}\ \emph {et~al.}(2011)\citenamefont {Blas},
  \citenamefont {Lesgourgues},\ and\ \citenamefont {Tram}}]{Blas:2011rf}%
  \BibitemOpen
  \bibfield  {author} {\bibinfo {author} {\bibfnamefont {D.}~\bibnamefont
  {Blas}}, \bibinfo {author} {\bibfnamefont {J.}~\bibnamefont {Lesgourgues}}, \
  and\ \bibinfo {author} {\bibfnamefont {T.}~\bibnamefont {Tram}},\ }\href
  {\doibase 10.1088/1475-7516/2011/07/034} {\bibfield  {journal} {\bibinfo
  {journal} {JCAP}\ }\textbf {\bibinfo {volume} {07}},\ \bibinfo {pages} {034}
  (\bibinfo {year} {2011})},\ \Eprint {http://arxiv.org/abs/1104.2933}
  {arXiv:1104.2933 [astro-ph.CO]} \BibitemShut {NoStop}%
\bibitem [{\citenamefont {Brinckmann}\ and\ \citenamefont
  {Lesgourgues}(2019)}]{Brinckmann:2018cvx}%
  \BibitemOpen
  \bibfield  {author} {\bibinfo {author} {\bibfnamefont {T.}~\bibnamefont
  {Brinckmann}}\ and\ \bibinfo {author} {\bibfnamefont {J.}~\bibnamefont
  {Lesgourgues}},\ }\href {\doibase 10.1016/j.dark.2018.100260} {\bibfield
  {journal} {\bibinfo  {journal} {Phys. Dark Univ.}\ }\textbf {\bibinfo
  {volume} {24}},\ \bibinfo {pages} {100260} (\bibinfo {year} {2019})},\
  \Eprint {http://arxiv.org/abs/1804.07261} {arXiv:1804.07261 [astro-ph.CO]}
  \BibitemShut {NoStop}%
\bibitem [{\citenamefont {Audren}\ \emph {et~al.}(2013)\citenamefont {Audren},
  \citenamefont {Lesgourgues}, \citenamefont {Benabed},\ and\ \citenamefont
  {Prunet}}]{Audren:2012wb}%
  \BibitemOpen
  \bibfield  {author} {\bibinfo {author} {\bibfnamefont {B.}~\bibnamefont
  {Audren}}, \bibinfo {author} {\bibfnamefont {J.}~\bibnamefont {Lesgourgues}},
  \bibinfo {author} {\bibfnamefont {K.}~\bibnamefont {Benabed}}, \ and\
  \bibinfo {author} {\bibfnamefont {S.}~\bibnamefont {Prunet}},\ }\href
  {\doibase 10.1088/1475-7516/2013/02/001} {\bibfield  {journal} {\bibinfo
  {journal} {JCAP}\ }\textbf {\bibinfo {volume} {02}},\ \bibinfo {pages} {001}
  (\bibinfo {year} {2013})},\ \Eprint {http://arxiv.org/abs/1210.7183}
  {arXiv:1210.7183 [astro-ph.CO]} \BibitemShut {NoStop}%
\bibitem [{\citenamefont {Lewis}(2019)}]{Lewis:2019xzd}%
  \BibitemOpen
  \bibfield  {author} {\bibinfo {author} {\bibfnamefont {A.}~\bibnamefont
  {Lewis}},\ }\href@noop {} {\  (\bibinfo {year} {2019})},\ \Eprint
  {http://arxiv.org/abs/1910.13970} {arXiv:1910.13970 [astro-ph.IM]}
  \BibitemShut {NoStop}%
\bibitem [{\citenamefont {Liddle}(2007)}]{Liddle:2007fy}%
  \BibitemOpen
  \bibfield  {author} {\bibinfo {author} {\bibfnamefont {A.~R.}\ \bibnamefont
  {Liddle}},\ }\href {\doibase 10.1111/j.1745-3933.2007.00306.x} {\bibfield
  {journal} {\bibinfo  {journal} {Mon. Not. Roy. Astron. Soc.}\ }\textbf
  {\bibinfo {volume} {377}},\ \bibinfo {pages} {L74} (\bibinfo {year}
  {2007})},\ \Eprint {http://arxiv.org/abs/astro-ph/0701113}
  {arXiv:astro-ph/0701113} \BibitemShut {NoStop}%
\bibitem [{\citenamefont {Ade}\ \emph {et~al.}(2016)\citenamefont {Ade} \emph
  {et~al.}}]{Planck:2015sxf}%
  \BibitemOpen
  \bibfield  {author} {\bibinfo {author} {\bibfnamefont {P.~A.~R.}\
  \bibnamefont {Ade}} \emph {et~al.} (\bibinfo {collaboration} {Planck}),\
  }\href {\doibase 10.1051/0004-6361/201525898} {\bibfield  {journal} {\bibinfo
   {journal} {Astron. Astrophys.}\ }\textbf {\bibinfo {volume} {594}},\
  \bibinfo {pages} {A20} (\bibinfo {year} {2016})},\ \Eprint
  {http://arxiv.org/abs/1502.02114} {arXiv:1502.02114 [astro-ph.CO]}
  \BibitemShut {NoStop}%
\bibitem [{\citenamefont {Marsh}(2016)}]{Marsh:2015xka}%
  \BibitemOpen
  \bibfield  {author} {\bibinfo {author} {\bibfnamefont {D.~J.~E.}\
  \bibnamefont {Marsh}},\ }\href {\doibase 10.1016/j.physrep.2016.06.005}
  {\bibfield  {journal} {\bibinfo  {journal} {Phys. Rept.}\ }\textbf {\bibinfo
  {volume} {643}},\ \bibinfo {pages} {1} (\bibinfo {year} {2016})},\ \Eprint
  {http://arxiv.org/abs/1510.07633} {arXiv:1510.07633 [astro-ph.CO]}
  \BibitemShut {NoStop}%
\bibitem [{\citenamefont {Antoniadis}\ \emph {et~al.}(2019)\citenamefont
  {Antoniadis}, \citenamefont {Karam}, \citenamefont {Lykkas}, \citenamefont
  {Pappas},\ and\ \citenamefont {Tamvakis}}]{Antoniadis:2018yfq}%
  \BibitemOpen
  \bibfield  {author} {\bibinfo {author} {\bibfnamefont {I.}~\bibnamefont
  {Antoniadis}}, \bibinfo {author} {\bibfnamefont {A.}~\bibnamefont {Karam}},
  \bibinfo {author} {\bibfnamefont {A.}~\bibnamefont {Lykkas}}, \bibinfo
  {author} {\bibfnamefont {T.}~\bibnamefont {Pappas}}, \ and\ \bibinfo {author}
  {\bibfnamefont {K.}~\bibnamefont {Tamvakis}},\ }\href {\doibase
  10.1088/1475-7516/2019/03/005} {\bibfield  {journal} {\bibinfo  {journal}
  {JCAP}\ }\textbf {\bibinfo {volume} {03}},\ \bibinfo {pages} {005} (\bibinfo
  {year} {2019})},\ \Eprint {http://arxiv.org/abs/1812.00847} {arXiv:1812.00847
  [gr-qc]} \BibitemShut {NoStop}%
\bibitem [{\citenamefont {AlHallak}\ \emph {et~al.}(2023)\citenamefont
  {AlHallak}, \citenamefont {Said}, \citenamefont {Chamoun},\ and\
  \citenamefont {El-Daher}}]{AlHallak:2022haa}%
  \BibitemOpen
  \bibfield  {author} {\bibinfo {author} {\bibfnamefont {M.}~\bibnamefont
  {AlHallak}}, \bibinfo {author} {\bibfnamefont {K.~K.~A.}\ \bibnamefont
  {Said}}, \bibinfo {author} {\bibfnamefont {N.}~\bibnamefont {Chamoun}}, \
  and\ \bibinfo {author} {\bibfnamefont {M.~S.}\ \bibnamefont {El-Daher}},\
  }\href {\doibase 10.3390/universe9020080} {\bibfield  {journal} {\bibinfo
  {journal} {Universe}\ }\textbf {\bibinfo {volume} {9}},\ \bibinfo {pages}
  {80} (\bibinfo {year} {2023})},\ \Eprint {http://arxiv.org/abs/2211.07775}
  {arXiv:2211.07775 [gr-qc]} \BibitemShut {NoStop}%
\bibitem [{\citenamefont {AlHallak}\ \emph {et~al.}(2022)\citenamefont
  {AlHallak}, \citenamefont {Chamoun},\ and\ \citenamefont
  {Eldaher}}]{AlHallak:2022gbv}%
  \BibitemOpen
  \bibfield  {author} {\bibinfo {author} {\bibfnamefont {M.}~\bibnamefont
  {AlHallak}}, \bibinfo {author} {\bibfnamefont {N.}~\bibnamefont {Chamoun}}, \
  and\ \bibinfo {author} {\bibfnamefont {M.~S.}\ \bibnamefont {Eldaher}},\
  }\href {\doibase 10.1088/1475-7516/2022/10/001} {\bibfield  {journal}
  {\bibinfo  {journal} {JCAP}\ }\textbf {\bibinfo {volume} {10}},\ \bibinfo
  {pages} {001} (\bibinfo {year} {2022})},\ \Eprint
  {http://arxiv.org/abs/2202.01002} {arXiv:2202.01002 [astro-ph.CO]}
  \BibitemShut {NoStop}%
\bibitem [{\citenamefont {Montefalcone}\ \emph {et~al.}(2023)\citenamefont
  {Montefalcone}, \citenamefont {Aragam}, \citenamefont {Visinelli},\ and\
  \citenamefont {Freese}}]{Montefalcone:2022jfw}%
  \BibitemOpen
  \bibfield  {author} {\bibinfo {author} {\bibfnamefont {G.}~\bibnamefont
  {Montefalcone}}, \bibinfo {author} {\bibfnamefont {V.}~\bibnamefont
  {Aragam}}, \bibinfo {author} {\bibfnamefont {L.}~\bibnamefont {Visinelli}}, \
  and\ \bibinfo {author} {\bibfnamefont {K.}~\bibnamefont {Freese}},\ }\href
  {\doibase 10.1088/1475-7516/2023/03/002} {\bibfield  {journal} {\bibinfo
  {journal} {JCAP}\ }\textbf {\bibinfo {volume} {03}},\ \bibinfo {pages} {002}
  (\bibinfo {year} {2023})},\ \Eprint {http://arxiv.org/abs/2212.04482}
  {arXiv:2212.04482 [gr-qc]} \BibitemShut {NoStop}%
\end{thebibliography}%

\end{document}